\documentstyle{laa}

\newcommand{\fig}[2] {\picplace{#2 cm}}

\newcommand{\proj}[1] {{#1'}}

\newcommand{\conv}[1] {{\overline{#1}}}
\newcommand{\obs}[1] {{#1_{\mbox{obs}}}}
\newcommand{\ave}[1] {\langle{#1}\rangle}
\newcommand{\ten}[1] {$10 ^{#1}$}
\newcommand{\ML}[1] {$\mbox{M/L}_{\mbox{#1}}$}
\newcommand{\kms} {$\mbox{km.s}^{-1}$}
\newcommand{\Msun} {$\mbox{M}_{\sun}$}
\newcommand{\Lsun} {$\mbox{L}_{\sun}$}
\newcommand{\Jansky} {$\mbox{J}_{\mbox{y}}$ }
\newcommand{\micron} {$\mu m$ }
\newcommand{\Kelvin} {\degr K }
\newcommand{\magsec} {$\mbox{mag.arcsec}^{-2}$} 
\newcommand{\Ag} {\AA~} 
\newcommand{\Wm} {$\mbox{W.m}^{-2}$} 
\newcommand{\diff} {\mbox{d}}
\newcommand{\dint} {\int\!\!\int}
\newcommand{\tint} {\int\!\!\int\!\!\int}
\newcommand{\Halpha} {$\mbox{H}_\alpha$}

\begin{document}

\thesaurus{11(11.09.1 M31; 11.11.1; 11.14.1; 11.16.1)}

\title{ Sub-arcsecond 2D photometry and spectrography of the nucleus
of M31:
the supermassive black hole revisited
\thanks{Based on observations taken with the Canada-France-Hawaii
Telescope, operated by
the National Research Council of Canada, the Centre National de la
Recherche Scientifique of France, and the University of Hawaii}}

\author{R. Bacon \inst{1}
\and E. Emsellem \inst{1}
\and G. Monnet \inst{2}
\and J.L. Nieto $^{\dag}$}

\offprints{R. Bacon (email: bacon@obs.univ-lyon1.fr)}

\institute{Observatoire de Lyon,
9 Avenue Charles Andr\'e, 69561 Saint Genis-Laval Cedex France
\and Canada-France-Hawaii Telescope Corporation, P.O Box 1597, Kamuela,
Hawaii 96743, USA}

\date{Received July 19; accepted August 13, 1993}

\maketitle

\begin{abstract}
Sub-arcsecond imagery (HRCAM, 0\arcsec.35 - 0\arcsec.57  FWHM)
and two-dimensional spectrography (TIGER, 0\arcsec.9 FWHM)
of the central nucleus of M31 have been obtained at
CFHT. The photometric data clearly show the double-peaked nucleus,
in excellent agreement with a recent HST image by Lauer et al. 1993.
We built deconvolved surface brightness models, using the
multi-Gaussian expansion method. We then perform a detailed
morphological analysis of the three central photometric
components (bulge, nucleus and bright secondary peak) and
derive various spatial luminosity models (oblate and triaxial).
Stellar velocity and velocity dispersion fields were derived
from the TIGER data: the former displays an
extremely rapid rotation around the true center of the galaxy,
while the latter exhibits a peaked structure offset
in the opposite direction of the brightest light peak.
Neglecting these offsets, both
extended versions of the virial theorem and detailed hydrodynamical
models confirm the classical strong central mass concentration,
perhaps a supermassive black hole of about 7.\ten{7} \Msun,
as well as a large excess of circular motions. These offsets, however,
suggest that the nucleus presently undergoes a strong stellar
oscillation, with non-stationary, non-axisymmetric dynamics.
This too reopens the case for a strong central mass concentration.

\keywords{galaxies: M31 --
          galaxies: nuclei of --
          galaxies: kinematic and dynamics of --
          galaxies: photometry and spectroscopy}
\end{abstract}

This paper is dedicated to our friend and collaborator Jean-Luc Nieto,
who met an untimely death in January 1992.

\section{Introduction}
Because of its proximity, M31 offers the best opportunity for a
high spatial resolution glimpse in the visible and near infra-red
spectral domains at the central region of a massive galaxy.

The existence of a highly contrasted central nucleus has been known for
a long time from visual observations, and has been dramatically
confirmed by the 1974 Light et al.'s Stratoscope~II images, with
the exceptional (even now!) spatial resolution of $\sim$ 0.2
arcsec. FWHM and despite a rather poor signal-to-noise ratio.

Moreover, Lallemand et al. (1960), with a then remarkable spatial
resolution of $\sim$ 1 arcsec., find spectroscopically that the
nucleus is rotating rapidly, exhibiting a very compact velocity curve,
returning basically to 0 \kms at a radius of 2 arcsec. Modern long-slit
spectroscopy, with much better signal to noise ratio, by
Kormendy (1988, hereafter K88), and Dressler and Richstone (1988,
hereafter DR88), at equal or slightly better spatial resolution,
fully confirmed this striking kinematic signature, and also
provided accurate velocity dispersion profiles.

Since the late 60s, it is known that the high velocity dispersion
and rotation in the nucleus of M31 imply either a large central
mass or a highly anisotropic kinematical regime. Detailed models
by K88 and DR88 have provided much evidence to imply a strong mass
concentration, of the order of a few \ten{7} solar masses, presumably
a supermassive black hole. It must be pointed out however, that
DR88, in contrast with their elaborate modelling (maximum entropy
technique), assume a spherical mass distribution though the
nucleus of M31 is known to be strongly flattened.
K88's approach uses a more realistic mass distribution, including a
constant flattening, but derives independently the mean velocity
and velocity dispersion curves, in a non self-consistent way.
Richstone et al. 1990, using basically the DR88 technique,
found indeed that the same experimental data can as well be fitted with
an extended unseen mass of $\sim$ \ten{8} \Msun, with a core radius
of $\sim$ 1.3 pc. On the other hand, Gerhard 1988 suggested that a
rotating nuclear bar, with no additional mass, could reproduce the
observations. This model, however, is only based on orbits analysis,
and is not self-consistent.  So far,
{\em there is no realistic self-gravitating model of M31's nucleus}.

The primary objective of this paper is to present new observations
with the aim of significantly improving our knowledge of M31's nucleus.
First high spatial
resolution (0\arcsec.35 - 0\arcsec.57 FWHM) high signal-to-noise ratio
photometry in B, R and I
bands has been obtained with HRCAM at CFHT, and
deconvolved, using the multi-Gaussian expansion technique (Monnet et al.
1992, Emsellem et al. 1993). They can now be compared with recent
V, I images from the HST (Lauer et al. 1993).
Sub-arcsecond (0.9 arcsec FWHM)
{\em two-dimensional} spectrographic observations
have been acquired with the TIGER spectrograph at CFHT, with the purpose
of highlighting the nucleus kinematical features (e.g. the displacement
of the kinematical center with respect to the maximum light location)
already glimpsed from long slit data.

The second objective of this paper is to present a realistic
self-gravitating model of the central part of M31. To this end
we again used the multi-Gaussian expansion technique
to build accurate spatial light distribution and
self-gravitating dynamical models.

The paper is organized as follows: in Sect. 2 we present the
photometric data as well as the corresponding light density models.
We then provide the kinematical data in Sect. 3. The first
part of Sect. 4 is devoted to discuss
the shape and mass distribution of the nucleus, while the
second part concentrates on
the observed asymmetries and their physical consequences.
We briefly conclude in Sect. 5. The two appendixes
deal with some particular aspects of the virial theorem.

\section{Photometry\protect\footnote{
The main observational data, including images and photometric
models are available through anonymous ftp at orion.univ-lyon1.fr
(134.214.4.6) in directory pub/m31. See file README for more information.}
}

\subsection{Observations}
The observations were made by Jean-Luc Ni\'eto at the prime focus of the
C.F.H. Telescope on November 7, 1991,
with the D.A.O. High Resolution Camera HRCAM.
This instrument is described in Mc Lure et al.
(1989). The detector was the RCA4 thin CCD, with 640x1024, $ 15 \mu $ pixels
and a $55 \mbox{e}^-\mbox{pixel}^{-1}$ r.m.s. read-out noise. During
that run, all images showed an unusually large number of low efficiency
columns,
probably due to a controller set-up problem.

The night was photometric, with superb ($\sim$0\arcsec.4 FWHM) seeing.
Several exposures were obtained in B, R and I (see
Table \ref{tab:obs1}), with a scale of 0.11 x 0.11 $\mbox{arcsec.}^2$ per
pixel and a total field of 70 x 113 arcsec. A bright star, located 45
arcsec. SW from the center, was used for fast tip-tilt
correction by the camera piezoelectric driven mirror.
%
\begin{table}
\begin{flushleft}
\begin{tabular}{rrrrr}
\hline
& Filter & Integ. & FWHM & FWHM \\
& Cousin System & Time & Major Axis & Minor Axis \\
\hline \hline
1 & B & 30 sec. & 0.60 \arcsec & 0.55 \arcsec \\
2 & R & 30 sec. & 0.60 \arcsec & 0.52 \arcsec \\
3 & I & 120 sec. & 0.37 \arcsec & 0.36 \arcsec \\
4 & I & 60 sec. & 0.37 \arcsec & 0.34 \arcsec \\
\hline
\end{tabular}
\end{flushleft}
\caption[]{HRCAM photometric observational parameters}
\label{tab:obs1}
\end{table}
%

\subsection{Data reduction}
A standard CCD reduction process (bias substraction and flatfielding) was
performed using the ESO MIDAS image processing package. The frames were
flux calibrated using the Johnson system and aperture photometry
available in the literature. No correction was applied to transform the
Cousin system to the Johnson system. The zero magnitude is accurate to 0.05
magnitude in all calibrated bands. No sky substraction was performed since
the sky background is negligible in the field of view (less than 1.5\%
and 0.5\% respectively at 10\arcsec and 3\arcsec from the center).
%
%

\subsection{Results}
The best resolved I band frame (exposure \#3) is presented in
Fig. \ref{fig:iso1}.
Major axis profiles (PA = 54\degr)
as well as ellipticities and  principal axes of the best fit ellipses
are given in Fig. \ref{fig:maj}, \ref{fig:ell} and \ref{fig:pa} for
this frame.
\begin{figure*}
\fig{fig1.ps}{16}
\caption[]{
Surface brightness isophotes of the I frame \#3 (central part).
Brightest isophote and isophote step are respectively 12.1 and
0.25 \magsec. North is 141\degr anti-clockwise from top.}
\label{fig:iso1}
\end{figure*}

As previously
observed by different authors (Light et al., 1974, Nieto et al., 1986,
Mould et al., 1989) the brightest light peak is offcenter by 0.5 arcsec NE
with respect to the center of the outer isophotes.

%
\begin{figure}
\fig{fig2.ps}{4.25}
\caption[]{
Major axis surface brightness I, R and B profiles. Solid lines are
for SW side and dashed lines for NE side.}
\label{fig:maj}
\end{figure}
%
\begin{figure}
\fig{fig3.ps}{4.25}
\caption[]{
Best fit ellipticities of I, R and B frames versus the semi-major
axis radius.}
\label{fig:ell}
\end{figure}
%
\begin{figure}
\fig{fig4.ps}{4.25}
\caption[]{
Best fit principal axis orientation of I, R and B frames versus the semi-major
axis radius.}
\label{fig:pa}
\end{figure}
%

We fully confirm that the nucleus appears as a quite distinct
entity from the bulge, with a significantly higher
ellipticity ($\epsilon_{\mbox{nucleus}} \simeq 0.4$,
$\epsilon_{\mbox{bulge}} \simeq 0.2$) and a different major axis position
angle ($ \mbox{PA}_{\mbox{nucleus}} \simeq 55 \degr$,
$\mbox{PA}_{\mbox{bulge}} \simeq 45 \degr $).
The outer parts ($r \ge 3$ arcsec.) present a very clumpy structure.
Images at different wavelengths are essentially the same, except
for small differences in the central region which can be
attributed to wavelength-dependent seeing.

A R-B color gradient image was computed after recentering and rebinning
of R and B frames. The medium spatial frequencies structures were
enhanced using an adaptive smoothing filter.
The resulting image (Fig. \ref{fig:color}) reveals two main absorbing
features already observed by Wirth et al. (1985):
\begin{itemize}
\item Two dust spiral arms starting from the nucleus and extending up to
the edge of the CCD frame.
\item A shell-like dust cloud (8 arcsec. diameter) located 16 arcsec. North
from the nucleus.
\end{itemize}
On the other hand, there is no evidence for dust extinction larger than
0.2 magnitude in the central 15 arcsec.; the slight color gradient,
that appears on Fig. \ref{fig:color} at the location of the very center,
is mostly due to the seeing and hence will not be further discussed.
%
\begin{figure}
\fig{fig5.ps}{10}
\caption[]{
Color gradient R-B frame showing the dust features in the central part
of M31. Coordinate orientation is similar to Fig. \ref{fig:iso1}.}
\label{fig:color}
\end{figure}
%

\subsection{Photometric Model}
\label{par:pproj}
We applied the Multi-Gaussian Expansion (MGE) technique to model the
central part of M 31 up to 40 arcsec. from the center using exposure
\#3 (I band). The bases of the
formalism can be found in Monnet et al. (1992), and a rather comprehensive
approach is given in Emsellem et al. (1993a, 1993b). The latter allows
non-centered light components,
as well as a Point Spread Function (hereafter PSF)
represented by an arbitrary sum of gaussian components.

Its main interest is to give a global approach to get qualitative and
quantitative information on the morphology of the galaxy. This method
provides also robust PSF deconvolution and line-of-sight deprojection.
The resulting 2D deconvolved surface brightness\footnote{
In this paper
we will use the following notation:
If $A$ denotes a spatial parameter, $\proj{A}$ indicates its line of sight
projected value, $\conv{A}$ its projected and PSF convolved value,
$\obs{A}$ its observed value and $\ave{A}$ its averaged value
over the projected surface.}
($\proj{\nu}$)
and 3D luminosity distribution model ($\nu$) are respectively given in the
following simple analytical form:
\begin{equation}
\proj{\nu} \left(x',y' \right) = \sum_i \proj{I}_i \exp{ \left\{
-\frac{1}{2 \proj{\sigma}_i^2} \left( {x'}_i^2 +
 \frac{{y'}_i^2}{\proj{q}_i^2} \right) \right\} }
\end{equation}
with
\begin{equation}
\left\{   \begin{array}{l}
x'_i = \left( x' - x'_{0_i} \right) \cos{ \left(\proj{\alpha}_i
\right) } + \left( y' - y'_{0_i} \right) \sin{
\left( \proj{\alpha}_i \right) } \\ y'_i = \left( x' - x'_{0_i} \right)
\sin{ \left(\proj{\alpha}_i \right) } - \left(
y' - y'_{0_i} \right) \cos{ \left( \proj{\alpha}_i \right) }
\\ \end{array}
\right.
\end{equation}
and
\begin{eqnarray}
\label{eq:nuspa}
\nu \left( x,y,z \right) & = & \sum_i I_i
\exp{ \left\{- \frac{1}{ 2 t_i u_i \sigma_i^2}
\left( t_i^2 \left( x - x_{0_i} \right)^2
\vphantom{ + u_i^2 \left( y - y_{0_i} \right)^2 +
\left( z - z_{0_i} \right)^2 } \right. \right. } \nonumber \\
&& \left. \left. \vphantom{  - \frac{1}{ 2 t_i u_i \sigma_i^2}
t_i^2 \left( x - x_{0_i} \right)^2 }
+ u_i^2 \left( y - y_{0_i} \right)^2 +
\left( z - z_{0_i} \right)^2 \right)  \right\}
\end{eqnarray}

The PSF was estimated using a bright star\footnote{In HRCAM
the reference star used
for the fast tip-tilt correction is not imaged on the CCD and therefore
can not be used to measure the PSF}
located 35 arcsec. SW from the
nucleus. Masking this star, we fitted a sum of 2D gaussian functions to
the galaxy and computed the residual image. The PSF is then modelized by
fitting a sum of 2D Gaussians to the background corrected star image.
A sum of two Gaussians gives an excellent fit ($\chi^2=0.026$, Fig.
%
\ref{fig:star}).
\begin{figure}
\fig{fig6.ps}{4.97}
\caption[]{
Model of the PSF: Star (dashed lines) and two Gaussians MGE model (solid
lines).
Right panel: Surface brightness contours.
Faintest isophote and isophote step are
respectively 500 and 50 ADU. Coordinate orientation
is similar than in Fig. \ref{fig:iso1}.
Left panel: corresponding major-axis profiles}
\label{fig:star}
\end{figure}
%

The parameters of the PSF model are given in Table \ref{tab:star}.
Note that the PSF is not circular and has larger wings
than a single 2D Gaussian. In this analysis, we assumed that the PSF,
derived from  the star image, is still valid for the central part of M 31, the
only region where its effect is dramatic.
%
\begin{table}
\begin{flushleft}
\begin{tabular}{rrrrr}
\hline
& $\proj{I}$ &$\proj{\sigma}(\arcsec)$ &
$\proj{q}$ & $\proj{\alpha}$ (\degr) \\
\hline \hline
1 & 3.50 & 0.13 & 0.94 & 120.1\\
2 & 1.21 & 0.31 & 0.85 & 83.5 \\
\hline
\end{tabular}
\end{flushleft}
\caption[]{I frame PSF model parameters. $\proj{I}$ was
normalized such as $\dint \mbox{PSF} \diff \proj{x} \diff \proj{y} = 1$}
\label{tab:star}
\end{table}
%

Following the MGE technique we represent the deconvolved surface
brightness ($\proj{\nu}$) by a sum of 2D gaussian functions
($G_i$):
\begin{equation}
\proj{\nu}\left(x', y'\right) = \sum_i G_i
\end{equation}
Hence the convolved surface brightness ($\conv{\nu}$) is given by :
\begin{equation}
\conv{\nu}\left(x', y'\right) = \proj{\nu} \otimes \mbox{PSF} =
\sum_i G_i \otimes \sum_p G_p = \sum_i \sum_p G_{i,p}
\end{equation}
\begin{equation}
\mbox{if} \; \mbox{PSF}\left(x', y'\right) = \sum_p G_p
\end{equation}
With $\obs{I}$, the observed surface brightness, minimizing :
\begin{equation}
\sum_{\mbox{pixels}} \left( \sum_i \sum_p G_{i,p}
\left(x', y'\right) - \obs{I}\left(x', y'\right)
\right)^2
\end{equation}
gives the deconvolved parameters $\proj{I}_i, \proj{x}_{0_i},
\proj{y}_{0_i},
\proj{\sigma}_i, \proj{q}_i, \proj{\alpha}_i$ of the $G_i$ functions.
The quality of the fit is measured by its corresponding $\chi^2$ value :
\begin{equation}
\chi^2 = \frac{\displaystyle{\sum_{\mbox{pixels}} \left( \sum_i
\sum_p G_{i,p} \left(x', y'\right)  - \obs{I}\left(x', y'\right)
\right)^2}}
{\displaystyle{\sum_{\mbox{pixels}}\obs{I}^2\left(x', y'\right)}}
\end{equation}

\subsubsection{Unconstrained fit (model U)}
This fit is obtained by letting all Gaussians parameters free of
constraints. We started with two Gaussians and
added new components until no more significant improvement of the
$\chi^2$ is obtained. An excellent fit was obtained with 11
Gaussians giving  $\chi^2$ = 1.7 \ten{-3}. The parameters of the model are
given in Table \ref{tab:unconst} and the deconvolved model is shown
in Fig. \ref{fig:deconv}. Note that two Gaussians of Table
\ref{tab:unconst} have negative intensities, emphasizing the fact
that each Gaussian is used only as a representative mathematical
function.
%
\begin{table}
\begin{flushleft}
\begin{tabular}{rrrrrrr}
\hline
& $\proj{x}_0$ (\arcsec) & $\proj{y}_0$(\arcsec) & $\proj{I}
(L_{\sun}.pc^{-2})$
& $\proj{\sigma}(\arcsec)$ & $\proj{q}$ & $\proj{\alpha}$ (\degr) \\
\hline \hline
1 &  0.01 &  0.11 & -2.029 \ten{4} &    0.175 & 0.602 & 151.54 \\
2 &  0.27 & -0.22 & 6.390 \ten{4} &    0.215 & 0.751 & 60.29 \\
3 & -0.29 &  0.34 & 1.476 \ten{5} &    0.231 & 0.735 & 74.47 \\
4 &  0.06 & -0.13 & -1.222 \ten{5} &    0.413 & 0.922 & 75.13 \\
5 &  0.00 & -0.00 & 2.292 \ten{5} &    0.576 & 0.688 & 51.78 \\
6 & -0.11 &  0.01 & 5.151 \ten{4} &    1.203 & 0.667 & 53.97 \\
7 &  0.68 &  0.17 & 5.603 \ten{3} &    1.815 & 0.733 & 39.40 \\
8 & -0.02 & -0.20 & 1.676 \ten{4} &    2.326 & 0.955 & 66.52 \\
9 &  0.04 &  0.02 & 1.143 \ten{4} &    6.272 & 0.786 & 43.68 \\
10 & -0.41 & -0.53 & 1.188 \ten{4} &   16.607 & 0.843 & 39.93 \\
11 &  0.85 & -0.05 & 1.568 \ten{4} &   59.680 & 0.900 & 49.32 \\
\hline
\end{tabular}
\end{flushleft}
\caption[]{Model U: MGE unconstrained fit parameters. $x_0$,
$y_0$  and $\proj{\alpha}$ are given in celestial coordinates.}
\label{tab:unconst}
\end{table}
%

\subsubsection{Constrained fit (model C)}
\label{par:const}
%
\begin{figure*}
\fig{fig7.ps}{14.3}
\caption[]{
Model C:
constrained 8 Gaussians fit (solid lines) of the I frame (dashed
lines).
Coordinate orientation is similar than in Fig. \ref{fig:iso1}.
Upper left panel: external part. Isophote step is 0.25 \magsec and
faintest isophote is 15 \magsec.
Upper right panel: Central part. Brightest isophote and isophote step are
respectively 12.1 and 0.25 \magsec.
Lower panel: major-axis profiles taken at $\proj{\alpha}$ = 43\degr :
outer part (left panel) and central part (right panel)}
\label{fig:fit}
\end{figure*}
%
\begin{figure}
\fig{fig8.ps}{16}
\caption[]{
Deconvolved images.
Upper panel (U): Unconstrained model. Lower panel (C): Constrained model.
Faintest isophote and
isophote step are respectively 14.1 and 0.25 \magsec.
Coordinate orientation is similar than in Fig. \ref{fig:iso1}}
\label{fig:deconv}
\end{figure}
%
Considering its large number (66) of free parameters, the previous fit
is obviously not an unique solution. Although no physical meaning can be
generally attached to each individual gaussian,
it appears reasonable to correlate each distinct
physical sub-component with a sub-set of one or more Gaussians. A quick
observation of the images shows that the central part of M31 may be split
in three components: the underlying bulge, the nucleus and the bright
peak. The bulge and nucleus are concentric while the brigth
peak is off-centered. We assume that each sub-component is axi-symmetric
and can then be fitted with concentric and aligned Gaussians.
We then apply the same procedure than for the unconstrained fit (see
previous paragraph); a total of 8
Gaussians (1 for the bright peak, 3 for the nucleus and 4 for the  bulge)
gives a good fit with $\chi^2$ (1.8 \ten{-3}) being only 4\% higher than
in the unconstrained fit. The final fit is shown in Fig. \ref{fig:fit}
and the deconvolved model in Fig. \ref{fig:deconv}.
Its Gaussians parameters are presented in Table \ref{tab:const}.
%
\begin{table}
\begin{flushleft}
\begin{tabular}{rrrrrrr}
\hline
& $\proj{x}_0$ (\arcsec) & $\proj{y}_0$(\arcsec) & $\proj{I}
(L_{\sun}.pc^{-2})$
& $\proj{\sigma}(\arcsec)$ & $\proj{q}$ & $\proj{\alpha}$ (\degr) \\
\hline \hline
\multicolumn{7}{c}{Bright Peak} \\
\hline
1 & -0.31 & 0.39 & 1.404 $10^5$ & 0.299 & 0.695 & 103.62 \\
\hline
\multicolumn{7}{c}{Nucleus} \\
\hline
1 & 0 & 0 & 1.054 $10^4$ & 0.595 & 0.452 & 54.01 \\
2 & 0 & 0 & 7.811 $10^4$ & 0.917 & 0.695 & 54.01 \\
3 & 0 & 0 & 1.738 $10^4$ & 1.728 & 0.658 & 54.01 \\
\hline
\multicolumn{7}{c}{Bulge} \\
\hline
1 & 0 & 0 & 1.650 $10^4$ & 2.268 & 1.000 & 42.75 \\
2 & 0 & 0 & 1.149 $10^4$ & 6.315 & 0.784 & 42.75 \\
3 & 0 & 0 & 1.182 $10^4$ & 16.639 & 0.854 & 42.75 \\
4 & 0 & 0 & 1.563 $10^4$ & 60.148 & 0.881 & 42.75 \\
\hline
\end{tabular}
\end{flushleft}
\caption[]{Model C: MGE constrained fit parameters}
\label{tab:const}
\end{table}
%

\subsubsection{Validity of the photometric deconvolved parameters}
\label{par:validity}
As for every deconvolution process, the MGE method can be strongly
affected by undersampling, and becomes unreliable for structures
not significantly larger than the PSF.
Tests of this method by Emsellem et al. (1993a) show that
uncertainties on the deconvolved surface brightness model
remain small for gaussian components at
least 1.5 times wider than the core of the PSF,
provided that the PSF is correctly sampled.

In our case, these two conditions are well met: the PSF
is well sampled (FWHM = 3.5 pixels) and the smallest deconvolved
gaussian width in Table \ref{tab:const} is almost two times
larger than the core of the PSF
($\frac{\sigma}{\sigma_{\mbox{psf}}} \geq 1.8$).

Our approach requires that the PSF, as extracted from a single star
in the field, remains valid at
the nucleus location. However, measurements of deviations from
isoplanetism on HRCAM images by McClure et al. (1991) have shown
variations of the image width as great as 30\% in the
100 arcsec. CCD field. From their Fig. 5, and taking into
account the 35 arcsec. separation
between the reference star and the nucleus, we can infer a
probable deterioration of the PSF of the order of 10\%.
To quantify its effect, we applied the MGE deconvolution again
using the PSF of Table \ref{tab:star}, broadened by 10\%. We found
no significant changes in the deconvolved gaussian parameters; this
emphasizes that all fitted structures are truly resolved.

As a last check, we processed the second I frame (exposure \#4)
in the same manner
than the first one. Despite a rather different PSF (Table \ref{tab:obs1}),
the process converged to a nearly identical solution: relative
differences of the smallest gaussian component parameters are
$\Delta \sigma = 0.9\%$, $\Delta q = 1.2\%$, $\Delta \alpha = 25\%$.

\subsubsection{Sub-components photometry}
{}From the decomposition achieved in Sect. \ref{par:const} (model C),
we can derive the global properties of each
sub-component
\footnote{We adopted a distance of 0.7 Mpc for M31 ($1\arcsec \simeq
3.39\mbox{pc}$)}
(Table \ref{tab:global}) as well as their relative
brightness as a function of radius (Fig. \ref{fig:comp}).
Note that this decomposition is based only on geometrical assumptions
without any a priori assumption on the components photometric law.

Even taking into account the uncertainties in the PSF
(see Sect. \ref{par:validity}), the bright peak appears truly resolved.
Its surface brightness, computed from the residuals between
the I frame and the bulge plus nucleus MGE model,
is presented in Fig. \ref{fig:resid}.
%
\begin{figure}
\fig{fig9.ps}{5}
\caption[]{
Residual frame between the I frame and the bulge plus nucleus MGE model C.
Inserted panel: star used as PSF model.
Faintest isophote and isophote step are 10\% of the maximum intensity.
Coordinate orientation is similar than in Fig. \ref{fig:iso1}}
\label{fig:resid}
\end{figure}
%

Note that the photometric parameters of the bulge shown in parenthesis in
Table \ref{tab:global}, refer to the gaussian components detected in the 40
arcsec. field only, and gives thus obviously a lower limit of its true
flux.
%
\begin{table}
\begin{flushleft}
\begin{tabular}{rrrrrrr}
\hline
Component & $\mbox{m}_{\mbox{I}}$ & $\mbox{M}_{\mbox{I}}$ &
$\mbox{r}_c$
& $\Delta\mbox{r}$ & $\ave{\mbox{q}}$ & $\ave{\alpha}$  \\
\hline \hline
Bright Peak & 13.74 & -10.7 & 1.2 & 1.8 & 0.69 & 104 \\
Nucleus & 11.13 & -13.3 & 3.0 & 0 & 0.61 & 54 \\
Bulge & (4.28) & (-20.1) & 35.5 & 0 & 0.88 & 43 \\
\hline
\end{tabular}
\end{flushleft}
\caption[]{
Global properties of photometric sub-components.
$\mbox{m}_{\mbox{I}}$ and $\mbox{M}_{\mbox{I}}$ are the apparent and
absolute magnitude in the I band (no correction for galactic
absorption has been applied). $\mbox{r}_c$ is the major-axis core
radius in
parsec. $\Delta\mbox{r}$ (center offset) is given relatively to the
nucleus center in parcsec.
$\ave{\mbox{q}}$ and $\ave{\alpha}$ are respectively the mean axis
ratio and principal axis (in degree).}
\label{tab:global}
\end{table}
%
\begin{figure}
\fig{fig10.ps}{5.65}
\caption[]{
Major axis deconvolved profile (PA = 54\degr) of nucleus (solid line),
bright peak(long-dashed line) and bulge (short-dashed line)
sub-components.}
\label{fig:comp}
\end{figure}
%

\subsubsection{Axisymmetric fits (models A and E)}
Although the nucleus and bulge principal axis are significantly misaligned
(with P.A.s
differing by 11\degr), deriving the best axisymmetric fit for the sum of
these two systems will prove quite useful later, for dynamical evaluations.
We have therefore substracted the gaussian component
corresponding to the bright peak from
the I image, and fitted the remaining surface brightness with a sum of
Gaussians, imposing the same center and the same principal axis.
Starting with the 7 components shown in Table \ref{tab:const}, we have
obtained an axisymmetric model with $\chi^2$ = 1.9 \ten{-3}.
Its parameters are given in Table \ref{tab:axi}.
%
\begin{table}
\begin{flushleft}
\begin{tabular}{rrrrrrr}
\hline
& $x_0$ (\arcsec) & $y_0$(\arcsec) & $\proj{I} (L_{\sun}.pc^{-2})$
& $\proj{\sigma}(\arcsec)$ & $\proj{q}$ & $\proj{\alpha}$ (\degr) \\
\hline \hline
1 & 0 & 0 & 1.059 $10^5$ & 0.595 & 0.454 & 52.75 \\
2 & 0 & 0 & 7.793 $10^4$ & 0.922 & 0.695 & 52.75 \\
3 & 0 & 0 & 1.732 $10^4$ & 1.731 & 0.661 & 52.75 \\
4 & 0 & 0 & 1.637 $10^4$ & 2.291 & 1.000 & 52.75 \\
5 & 0 & 0 & 1.153 $10^4$ & 6.346 & 0.795 & 52.75 \\
6 & 0 & 0 & 1.185 $10^4$ & 16.728 & 0.874 & 52.75 \\
7 & 0 & 0 & 1.544 $10^4$ & 61.061 & 0.892 & 52.75 \\
\hline
\end{tabular}
\end{flushleft}
\caption[]{Model A: MGE axisymmetric nucleus and bulge fit
parameters}
\label{tab:axi}
\end{table}
%

In order to apply the tensor virial theorem
to derive the global mass distribution, we will need later
(Sect. \ref{par:vir}) an even more constrained model, in which all
Gaussians of each individual system (i.e. nucleus and bulge)
share the same axis ratio and principal axes.
Parameters of this model are given in Table \ref{tab:vir}
($\chi^2$ = 1.8 \ten{-3}).
%
\begin{table}
\begin{flushleft}
\begin{tabular}{rrrrrrr}
\hline
& $x_0$ (\arcsec) & $y_0$(\arcsec) & $\proj{I} (L_{\sun}.pc^{-2})$
& $\proj{\sigma}(\arcsec)$ & $\proj{q}$ & $\proj{\alpha}$ (\degr) \\
\hline \hline
\multicolumn{7}{c}{Nucleus} \\
\hline
1 & 0 & 0 & 8.697 $10^4$ & 0.518 & 0.610 & 53.92 \\
2 & 0 & 0 & 7.696 $10^4$ & 0.880 & 0.610 & 53.92 \\
3 & 0 & 0 & 2.132 $10^4$ & 1.394 & 0.610 & 53.92 \\
\hline
\multicolumn{7}{c}{Bulge} \\
\hline
1 & 0 & 0 & 2.332 $10^4$ & 2.012 & 0.877 & 44.54 \\
2 & 0 & 0 & 1.246 $10^4$ & 5.447 & 0.877 & 44.54 \\
3 & 0 & 0 & 1.218 $10^4$ & 15.531 & 0.877 & 44.54 \\
4 & 0 & 0 & 1.161 $10^4$ & 58.156 & 0.877 & 44.54 \\
\hline
\end{tabular}
\end{flushleft}
\caption[]{Model E: MGE elliptical fit parameters}
\label{tab:vir}
\end{table}
%

\subsubsection{Comparison between the various MGE models and HST data}
As illustrated by Fig. \ref{fig:pfit},
the unconstrained model (U)
gives the more accurate description of M31's central part
amongst the four MGE photometric models.
Although the
constrained model (C) is slightly less refined than the U model,
it allows a clearer separation
between physical sub-components. However one must recall
that this decomposition based on a few geometrical assumptions
(axisymmetry) is not unique. Different assumptions leading to
other models are obviously a possibility.  For instance, we
could have used
standard photometric (e.g. $r^{1/4}$ or exponential), which, however,
imply some ad-hoc hypothesis in addition to the above geometrical assumptions.

The last two models A and E,
should be taken only as approximations of M31's
central surface brightness. They are intended to be used in
the dynamical modelling which generally requires
additional geometrical simplifications.

%
\begin{figure}
\fig{fig11.ps}{5.65}
\caption[]{
Major axis residuals in magnitude of the different MGE models.
{}From top to bottom: models E, A, C and U.
Note that the bright peak gaussian component of the model C has
been included in models E and A for the comparison.
}
\label{fig:pfit}
\end{figure}
%

External comparison is given by the recent deconvolved HST V band image
of Lauer et al. (1993). According to the authors, the resolution of
this image is close to the nominal diffraction limit
but with some residual blurring within a 0.1 arcsec. radius.
Even in that case, the resolution is twice as good as
Stratoscope's data (Light et al. 1974).

As illustrated by Fig. \ref{fig:hst}, the spatial resolution of
the deconvolved U model is only slightly lower than HST deconvolved data.
Note that the deconvolved U model retrieves the double peak,
while the C model only presents an
inflexion point.
%
\begin{figure}
\fig{fig12.ps}{4.00}
\caption[]{
Comparison between Lauer et al. (1993) HST deconvolved data (V band) and
MGE deconvolved U and C model (I band).
Short dashed line: Cut at PA = 43\degr~ of HST V data (short dashed line)
from Lauer et al. (1993) and of the MGE deconvolved U model (long dashed
line) and C model (solid line). The level of HST V data has been normalized
($\mu_V - 1.85$) to fit the external part of our I band models.
The original HRCAM I band profile (dots) is given for comparison}
\label{fig:hst}
\end{figure}
%

Despite a somewhat lower resolution,
we found a remarkable agreement between
the photometric properties of the sub-components
inferred by Lauer et al. (1993) with
the ones derived from the C model:
\begin{itemize}
\item Separation between the bright peak and the nucleus center
(referred respectively as P1 and P2 in Lauer et al. 1993) is
$0.49 \pm 0.01$ arcsec.
according to the authors, versus 0.50 arcsec. in the C model.
\item Computed bright peak and nucleus core radii are nearly identical
(respectively 0.37 and 0.82 arcsec. with the HST
versus 0.35 and 0.88 arcsec. for the C model)
\item Geometric parameters are also comparable. According to the various
fit of the nucleus, Lauer et al., found an apparent axis ratio $\sim$ 0.6
- 0.7 and a PA $\sim$ 60\degr, while we found respectively
0.61 and 54\degr~ for the axis ratio and the principal
axis orientation.
The line joining the center of the nucleus and the bright peak has
a position angle of 38\degr, to be compared with the HST value of
43\degr. The measurement of the bright peak axis ratio and
principal axis give respectively 0.69 and $\simeq$ 104\degr,
consistent with the Figure 2 of Lauer at al. (1993).
\end{itemize}

\subsection{Space light density}
\label{par:pspac}
One of the main interest of the MGE technique, is to give analytical line
of sight  deprojection assuming that individual Gaussians have a triaxial
ellipsoidal shape
\footnote{Note that even if each individual gaussian component
is constant on homothetic ellipsoid, this is not generally
the case for their overall sum}.
Knowledge of the three Euler angles  of the line of view
is required to derive the unique solution of the
deprojection. Unfortunately, these angles are not generally known, and some
a priori assumptions are needed to retrieve the true shape.

Although deprojection is possible for both the constrained and
the unconstrained
fits, we will  consider only the constrained ones. In that cases,
assumptions concerning shapes and axis of symmetry are more easily
formulated. A number of different hypotheses are
explored in the following sub-sections.

\subsubsection{Axisymmetric model}
If we assume that each Gaussian has an axisymetric (oblate) shape,
only one Euler angle
(the inclination $\theta$ of the line of sight) is needed to deproject
the light model.
In the case of an axisymmetric fit, it is natural to assume the same
inclination $\theta =  77\degr$ for the nucleus and the bulge as for
the M31's main disk. The resulting MGE spatial parameters are given in Table
\ref{tab:axispace}
for the axisymmetric model A and in Table \ref{tab:ellspace}
for the elliptic model E.
%
\begin{table}
\begin{flushleft}
\begin{tabular}{rrrr}
\hline
& $I$ & $\sigma$(\arcsec) & t = u \\
\hline \hline
1 & 7.967 $10^4$ & 0.59 & 0.40 \\
2 & 3.473 $10^4$ & 0.92 & 0.67 \\
3 & 4.135 $10^3$ & 1.73 & 0.64 \\
4 & 2.850 $10^3$ & 2.29 & 1.00 \\
5 & 7.362 $10^2$ & 6.34 & 0.78 \\
6 & 2.849 $10^2$ & 16.73 & 0.87 \\
7 & 1.016 $10^2$ & 61.06 & 0.89 \\
\hline
\end{tabular}
\end{flushleft}
\caption[]{MGE spatial parameters
(model A). $I$ is in $\mbox{L}_{\sun}\mbox{pc}^{-2}(\arcsec)^{-1}$}
\label{tab:axispace}
\end{table}
%
\begin{table}
\begin{flushleft}
\begin{tabular}{rrrr}
\hline
& $I$ & $\sigma$(\arcsec) & t = u \\
\hline \hline
\multicolumn{4}{c}{Nucleus} \\
\hline
1 & 7.026 $10^4$ & 0.52 & 0.58 \\
2 & 3.657 $10^4$ & 0.88 & 0.58 \\
3 & 6.397 $10^3$ & 1.39 & 0.58 \\
\hline
\multicolumn{4}{c}{Bulge} \\
\hline
1 & 4.662 $10^3$ & 2.01 & 0.87 \\
2 & 9.201 $10^2$ & 5.45 & 0.87 \\
3 & 3.155 $10^2$ & 15.53 & 0.87 \\
4 & 1.116 $10^2$ & 58.15 & 0.87 \\
\hline
\end{tabular}
\end{flushleft}
\caption[]{MGE spatial parameters
(model E). $I$ is in $\mbox{L}_{\sun}\mbox{pc}^{-2}(\arcsec)^{-1}$.
Note that the spatial orientation of the nucleus and the bulge must
be compatible with a tilt of 11\degr.}
\label{tab:ellspace}
\end{table}
%

\subsubsection{Triaxial model}
Following Stark (1977) and Nieto (1984), we will assume that the nucleus
shares a common equatorial plane with the bulge and the disk. In that
case the observed tilt between the nucleus and the disk (21\degr) and
between the bulge and the disk (9.75\degr) would be solely due to a projection
effect. Two Euler
angles are then fixed ($\theta = 77\degr$, $\psi =
21\degr$ for the nucleus and $\theta = 77\degr$, $\psi = 9.75\degr$
for the bulge), the third one
($\phi$m) beeing restricted to a rather small range
[103\degr  - 127\degr ] by the smaller axis ratio of the Gaussian nucleus
component ($q = 0.45$). The parameters of the intermediate model
($\phim = 115\degr$) are given in Table \ref{tab:trispace}.
%
\begin{table}
\begin{flushleft}
\begin{tabular}{rrrrrrrr}
\hline
& $\phi$(\degr) & $\theta$(\degr) & $\psi$(\degr)
& $I$ & $\sigma$(\arcsec) & t & u \\
\hline \hline
\multicolumn{8}{c}{Nucleus} \\
\hline
1 & 115 & 77 & 21
  & 7.111 $10^4$ & 0.60 & 0.22 & 0.70 \\
2 & 115 & 77 & 21
  & 2.854 $10^4$ & 0.92 & 0.42 & 0.89 \\
3 & 115 & 77 & 21
  & 3.419 $10^3$ & 1.73 & 0.39 & 0.87 \\
\hline
\multicolumn{8}{c}{Bulge} \\
\hline
1 & 115 & 77 & 9.75
  & 2.901 $10^3$ & 2.27 & 1.00 & 1.00 \\
2 & 115 & 77 & 9.75
  & 7.820 $10^2$ & 6.32 & 0.84 & 0.76 \\
3 & 115 & 77 & 9.75
  & 2.979 $10^2$ & 16.64 & 0.90 & 0.84 \\
4 & 115 & 77 & 9.75
  & 1.080 $10^2$ & 60.15 & 0.92 & 0.87 \\
\hline
\end{tabular}
\end{flushleft}
\caption[]{MGE spatial parameters (triaxial model).
$I$ is in $\mbox{L}_{\sun}\mbox{pc}^{-2}(\arcsec)^{-1}$}
\label{tab:trispace}
\end{table}
%

\subsubsection{Alternative models}
We could as well deproject separately each sub-system in model C
with different Euler angles. It is however difficult to favour any
particular set, so we just give an example where each system is
oblate and is seen with an inclination of 77\degr (Table \ref{tab:obspace}).
We assumed that the
spatial centers of the nucleus and the bulge are identical. The true spatial
location of the bright peak is unknown since we do not know its
coordinate along the line of sight.
%
\begin{table}
\begin{flushleft}
\begin{tabular}{rrrrrrr}
\hline
& $\mbox{PA}$(\degr) & $x_c$(\arcsec) & $y_c$(\arcsec)
& $I$ & $\sigma$(\arcsec) & t \\ \hline \hline
\multicolumn{7}{c}{Bright Peak} \\
\hline
1 & 50 & -0.48 & 0.14
  & 1.928 $10^5$ & 0.30 & 0.68 \\
\hline
\multicolumn{7}{c}{Nucleus} \\
\hline
1 & 0 & 0.00 & 0.00
  & 7.925 $10^4$ & 0.60 & 0.40 \\
2 & 0 & 0.00 & 0.00
  & 3.497 $10^4$ & 0.92 & 0.68 \\
3 & 0 & 0.00 & 0.00
  & 4.159 $10^3$ & 1.73 & 0.63 \\
\hline
\multicolumn{7}{c}{Bulge} \\
\hline
1 & -11 & 0.00 & 0.00
  & 2.901 $10^3$ & 2.27 & 1.00 \\
2 & -11 & 0.00 & 0.00
  & 7.383 $10^2$ & 6.32 & 0.77 \\
3 & -11 & 0.00 & 0.00
  & 2.863 $10^2$ & 16.64 & 0.84 \\
4 & -11 & 0.00 & 0.00
  & 1.045 $10^2$ & 60.15 & 0.87 \\
\hline
\end{tabular}
\end{flushleft}
\caption[]{MGE spatial parameters (alternative
model). PA is the rotation angle of each subcomponent
after projection with an inclination $\theta = 77$ \degr.
$I$ is in $\mbox{L}_{\sun}\mbox{pc}^{-2}(\arcsec)^{-1}$.}
\label{tab:obspace}
\end{table}
%
\subsubsection{Comparison of the spatial light density models}
We have seen in the previous sections some of the possible light
density models compatible with the observed photometry. Rather than
exploring the infinite number of deprojected models, we
choosed to bracket the various solutions, using simple geometrical
assumptions.

The axisymmetric model is the simplest one. However this model
is not able to reproduce the observed tilts between the nucleus,
the bulge and the disk. Furthermore, the bright peak is excluded.
It is nevertheless quite valuable for dynamical modelling purposes.

The triaxial model is more realistic because it takes into account
the observed tilts. In that models the three components share a common
principal plane, which is required for stability arguments. Again a
large number of solutions are possible, but all present an elongated
bar-like shape seen nearly end-on for the nucleus. Nevertheless, the bright
peak is again excluded.

If we want to take into account all the components, including the
bright peak, there are no simple geometrical assumptions
which could help us to reduce the number of free parameters. The last
model, given as an exemple, has no reason to be favored amongst
the others.
It is clear that this photometric analysis cannot by itself
determines the true 3D morphology of the nucleus of M31, hence
kinematical data and analysis are needed to go further.
These are developed in the following sections.

\section{Spectrography\protect\footnote{
As for photometric data, the main data, including
selected spectra and broadening functions, velocity and
velocity dispersion field and dynamical models
are available through anonymous ftp at orion.univ-lyon1.fr
(134.214.4.6) in directory pub/m31. See file README for more information.}
}
%
\subsection{The integral field spectrograph TIGER}
The Integral Field Spectrograph (hereafter IFS) TIGER is described
by Courtes et al.
(1987) and Bacon et al. (1988).  This spectrograph built by the
Observatoire de Lyon and the Observatoire de Marseille,
is a true 3D spectrograph which offers a complete 2D spatial mapping and low
to medium spectral resolution. It is attached at the f/8 Cassegrain
focus of the 3.6~m CFH telescope.

A focal enlarger gives a large-scale image of the field split up by
an array of 400 microlenses.  Each microlens ({\large\o}~=~1.4~mm)
corresponds to one spatial element of the final
monochromatic images and gives a very small image ({\large\o}~=~45~$\mu$)
of the pupil of the telescope. Contrary to an optical fiber mounting,
the lenses transmit light with negligible changes in the optical
throughput and etendue. This gives freedom to change the
spatial sampling from 0.3 to 0.6 arcsec. On the other hand the spectra
from the edge of the field are truncated.
The lens array is then followed by a classical spectrographic optical
design with a grism as the dispersor. The array is rotated by a few
degrees to avoid superimposition of the spectra in the direction
of the dispersion. The dispersed pupils are then imaged on a CCD
detector with a camera.

With the TIGER spectrograph, spatial resolution is fully determined
before the
microlenses array. Neither the collimator nor the camera aberrations
can degrade the image quality, as each spatial element is split up several
millimeters apart. This is not the case in long slit spectrography or
in multi-fiber spectrography when the fibers are densely packed along the
slit.

The superiority of the TIGER IFS upon the classical long-slit
techniques lies in its bidimensional spatial sampling. This allows
for example to
derive true velocity fields instead of velocity curves.
Furthermore the spatial sampling can be easily adapted to match the
seeing without the positioning problems of very narrow slits. Finally,
the seeing can be measured {\em a posteriori} from the reconstructed images
which will be decisive for comparison of the data with theoretical models.

\subsection{Observations}

The nucleus of M31 was observed in November 1990 and September 1991,
with spatial resolution ranging from 0.7 to 1 arcsec. FWHM. Due to
improvements of the instrument, the September 1991's data benefited
from a better spectral resolution than the November 1990 ones.
Apart from that, they are highly compatible, and we will mostly
restrict the discussion to the last data set, just pointing out any
similarity or discrepancy between the two sets, when warranted.

We used a 10 mm focal length enlarger which gives a spatial sampling
of 0.39 arcsec. per micro-lens and a field of view of 7.5x7.5
$\mbox{arcsec}^2$.
This is well adapted to the 0.7 arcsec. FWHM median image quality at CFHT.

A total of five 30 mn exposures were obtained in good seeing
conditions (0.7 arcsec. FWHM). The first three exposures were taken in
the 5040 \Ag - 5690 \Ag wavelength interval where the absorption
lines (MgII, Fe, Ca) can be used to derive stellar kinematics.
Table \ref{tab:obs2} gives the observational parameters of this
configuration. The
two remaining exposures were taken in the 6430 \Ag - 7110 \Ag
wavelength interval to look for possible ionized gas emission lines
([NII], H$\alpha$`, [SII]).
%
\begin{table}
\begin{flushleft}
\begin{tabular}{ll}
\hline
Spatial sampling & 0.39 \arcsec \\
Field of view & 7.5 \arcsec x 7.5 \arcsec \\
Number of spectra & 400 \\
Spectral sampling & 1.7 \Ag.$\mbox{pixel}^{-1}$ \\
Spectral resolution & 3.3 \Ag (FWHM) \\
Instrumental broadening & 101 \kms \\
Wavelength interval & 5100 - 5580 \Ag \\
Standard stars & $\eta$ Cygni (KOIII) \\
               & $\eta$ Ophiucus (KOIII) \\
               & 39 Cygni (K1III) \\
\hline
\end{tabular}
\end{flushleft}
\caption[]{TIGER spectrographic observational parameters}
\label{tab:obs2}
\end{table}

\subsubsection{Absorption lines}

The first exposure was centered on the nucleus and the two others were
shifted 3 arcsec. along the nucleus major axis. All exposures
overlapped in the center, so that the nuclear region is oversampled.
K0III stars were observed
to be used as templates for the spectral deconvolutions. To get a roughly
uniform illumination of the field we moved the telescope during the star
exposures.

The channeled spectrum of a continuum quartz lamp, shining through a highly
stable Fabry-Perot interferometer, was used for spectral calibrations
before and after each observation.
The advantage of such
a procedure is that calibration peaks given by the Fabry-Perot etalon
are regularly spaced and not blended.
This allows a much more
precise wavelength calibration over the whole spectral range
than with classical discharge lamps (Bacon et al, 1990).

\subsubsection{Emission lines}
The two 30 mn exposures were centered on the nucleus, with a small
shift of 0.8 arcsec. between them along the major-axis.
Spectral range and resolution
were respectively 6500 - 7000 \Ag and 1900. Some standard stars have
been observed to be used as flux calibrators.

\subsection{Data reduction}

A dedicated software has been developed (100 000 C lines code) to
reduce and analyze the TIGER data. This software uses the ESO MIDAS
package
as a display and image processing environment and the NAG library for
numerical analysis. Further details can be found in
Rousset et al (1992), and a new revised version of the reduction
methods will be described in a forthcoming paper.

\subsubsection{Preprocessing and data extraction}
\paragraph{Image processing}

A median of 20 bias exposures is computed. The mean value of the CCD
overclock window is normalized with respect to the corresponding mean
value in the object frame. A high pass filter applied to a median of
continuum lamp exposures provides the high frequency flatfield. The
observed pixel to pixel flatfield variations are small (a few percent)
with the SAIC1 CCD.  Each object frames is then bias subtracted and
flatfielded. Dark current ($< 5$ ADU) has been neglected for the 30 mn
exposures.

\paragraph{Spectra extraction}

We use a continuum lamp frame with high signal-to-noise ratio to determine an
accurate position (sub-pixel) for each spectrum.
This allows us to follow the curvature caused by optical distortions.
We take advantage of the fact that the
PSF is a pupil image (i.e. image of the primary mirror convolved with
the instrumental broadening function) to model it with an empirical
one-dimensional PSF, in the direction perpendicular to the dispersion.
This PSF is common to all exposures, and
each spectrum is extracted from the object and calibration frames
by fitting the intensity of the PSF model to the cross-dispersion
profiles. The micro-lenses position associated with a spectrum
gives the location of the spatial element and is saved in a table.
The three 30 mn exposures were processed individually, each one providing
400 spectra.

\paragraph{Wavelength calibration}

We first evaluate the interference orders of the Fabry-Perot
calibration peaks. This is done using a neon lamp reference exposure
obtained in the same conditions. Knowing the absolute wavelength
of the Fabry-Perot calibration peaks, we apply a classical wavelength
calibration process on the observed spectra.
The mean residual, after fitting third degree
polynomials, was 0.012 \Ag with a standard deviation of 0.006 \Ag only.

\paragraph{Low frequency flatfielding}

The wavelength calibrated spectra extracted from a sky or dome
exposure are used to remove low frequency flatfield variations. The
correction factor does not exceed 10\% except at the CCD
edges where the camera vignetting produces a 50\% flatfield correction.

\paragraph{Cosmic rays removal}

Some of the cosmic rays are removed during the spectra extraction
using a two pass sigma-kappa algorithm. The remaining ones are removed
using the comparison between each spectrum with the median of its
spatial neighbours at each wavelength.
The detected cosmic rays are then replaced by the corresponding
median values.

\subsubsection{Building images}

Physical quantities extracted from spectra are saved
as new columns in the table created by the spectra extracting
process. Using the two spatial coordinates of the micro-lens as
independent variables, images of physical quantities are computed
through triangulating and interpolating processes.

Images through simulated filters are then easily built from the
integrated
flux in the chosen wavelength range. These images thus give the precise
location of each spectrum. It is then, in principle, possible to measure
the true spatial resolution of the spectrographic data,
using the continuum image of a
point-like object in the field. Unfortunately no such object is present
in M31 TIGER's field, and we had to resort to a more indirect
way (see Sect. \ref{par:seeing}).

\subsubsection{Stellar velocity and velocity dispersion fields}

All spectra were first truncated to a common wavelength interval (5150
\Ag - 5430\Ag). Spectra which do not fill entirely this interval were
discarded. Each spectrum was rebinned in $\log(\lambda)$ and its continuum
removed using a fifth degree polynomial fit.

Standard star and galaxy spectra are processed exactly in the same
manner.
We then used the Fourier Correlation Quotient (FCQ) method of Bender (1990)
to derive the full line of sight velocity profile (broadening function).
This method is more powerfull
than the cross-correlation (Tonry and Davies 1979) or Fourier
quotient (Sargent et al, 1977) techniques because it does not
assume any a priori
shape for the broadening function. This avoids the
known pitfall of Gaussian fits which leads to systematic errors when different
kinematical subcomponents are mixed onto the line of sight (Bender
1992).

Mean velocity and velocity dispersion are derived from the first two moments
of the broadening function or by gaussian fitting.
The mean velocity and velocity dispersion fields are computed from the
individual values by triangulation and interpolation.
We use $\eta$ Cygni as the reference star to derive the broadening function
and check that other stars give consistent results.

\subsection{Results}

\subsubsection{Reconstructed image}
\label{par:seeing}
%
%
\begin{figure}
\fig{fig13.ps}{8.0}
\caption[]{
Reconstructed frame (5300 - 5400 \Ag) from the three merged exposures.
The horizontal axis has been aligned with the nucleus photometric
major axis (PA = 54\degr). Image has normal celestial handedness.
Location of the nucleus photometric center is marked with a white circle.
Maximum brightness isophote and isophote step are respectively
8.05 and 0.16 ($2.5 \log{I}$ unit).}
\label{fig:recon}
\end{figure}
Individual exposures have been independently processed. Reconstructed
continuum images were computed from the total flux in the  5300 \Ag - 5400
\Ag
range. The precise location of the brightest peak measured on these
images was used as a common reference for the three exposures.

The knowledge of the true spatial resolution achieved in the
spectrographic exposures is essential to the interpretation of the
stellar kinematics. The computation of this
parameter is more difficult here because of the lack of a point-like
source in the field.

The approach is to compare the reconstructed image
($\mbox{I}_{\mbox{obs}}$)
with a direct image ($\mbox{I}_{\mbox{ref}}$), obtained with a
significantly better spatial resolution (which, fortunately, is
generally the case).
Apart from coordinates translation and
rotation, differences between the two frames are due to
the convolution of the reference exposure with an unknown function.
We then minimize:
\begin{equation}
\sum_{\mbox{pixels}} \left[ k \mbox{I}_{\mbox{ref}}
\left(x,y\right) \otimes G\left(x,y\right) -
\mbox{I}_{\mbox{obs}}\left(x',y'\right) \right]^2
\end{equation}
with the coordinate translation and rotation:
\begin{equation}
\left\{   \begin{array}{l}
x' = \left( x - dx \right) \cos{\beta} + \left( y - dy \right)
\sin{\beta} \\
y' = -\left( x - dx \right) \sin{\beta} + \left( y - dy \right)
\cos{\beta}
\\ \end{array}
\right.
\end{equation}
and $k$ being a normalization factor.
The convolution function G is assumed to be a circular Gaussian:
\begin{equation}
G \left(x,y\right) = \exp \left(-\frac{\left(x^2 + y^2\right)}
{2 \sigma_0^2}\right)
\end{equation}
A non-linear least square minimization gives then
the 5 free parameters $dx$, $dy$ (translation), $\beta$ (rotation),
$k$ (normalization factor) and $\sigma_0$ (sigma of the gaussian
function).
Assuming that the PSF of the reference frame ($\sigma_{\mbox{ref}}$)
is also gaussian,
the absolute TIGER PSF ($\sigma_{\mbox{tiger}}$) is then simply:
\begin{equation}
\sigma_{\mbox{tiger}} = \sqrt{\sigma_{\mbox{ref}}^2 + \sigma_0^2}
\end{equation}
The fact that the PSF is not truly gaussian for both reference
and TIGER frames is responsible for the main error
in this derivation. The uncertainty on the derived
FWHM can be roughly estimated to be about 5\%.
This minimization provides also
the fine offcentering and the global relative
intensity between the three spectrographic exposures.

The three TIGER reconstructed images were then compared with the
HRCAM I band frame, neglecting the small color gradients between I
and V in the center.
Fits are very good ($\chi^2$ = 2. \ten{-4}):
isophotes of the transformed HRCAM I frame are nearly
indistinguishable from the reconstructed TIGER exposures. Output
parameters are given in Table \ref{tab:seeing}.
%
\begin{table}
\begin{flushleft}
\begin{tabular}{rrrrrrr}
\hline
& $\sigma_0$ & $dx$ & $dy$ & $\beta$ & $k$ & $\sigma_{\mbox{tiger}}$\\
\hline \hline
C & 0.31 & 0.24 & -0.14 & -137 &  3.997 \ten{-3} & 0.35 \\
NW & 0.33 & 0.27 & -0.13 & -138 &  4.075 \ten{-3} & 0.36 \\
SE & 0.36 & 0.26 & -0.16 & -141 &  3.975 \ten{-3} & 0.40 \\
\hline
M & 0.33 & 0.25 & -0.12 & -139 &   & 0.37 \\
\hline
\end{tabular}
\end{flushleft}
\caption[]{Comparison between the TIGER reconstructed
frame and the HRCAM I exposure ($\sigma_{\mbox{hrcam}}$ = 0\arcsec.16).
$\sigma_0$, $dx$, $dy$ and $\sigma_{\mbox{tiger}}$ are in arcsec.
$\beta$ is in degree. C (central), NW (North-West) and SE (South-East)
marked the
frame center with respect to the nucleus center. M is for the merged
frame}
\label{tab:seeing}
\end{table}

All images look very similar in flux ($\Delta k = 3\%$) and spatial
resolution ($\Delta \sigma_{\mbox{tiger}} = 14\%$).
The offset between a lens in the central exposure and the nearest
neighbouring lens in the SW (respectively NE) exposure is
0.08 arcsec. (respectively 0.23 arcsec.). These values
are small relatively to the individual spatial resolution, so we
decided to merge the three data set, averaging spectra when the distance
between the corresponding lenses is less than 0.3 arcsec.
Note that the average was weighted
by the global relative intensity between each frame and the HRCAM
image (parameter $k$). The location of the averaged spectra is then
the corresponding weighted center of gravity. The 620 spectra
of the merged data set covered a region of 11 x 9.5 $\mbox{arcsec.}^2$
with a high overlapping in the central 3 x 3 $\mbox{arcsec}^2$.
This procedure increases the signal to noise ratio (S/N $\sim$ 120
at the center) in the
overlapping region, without much affecting the spatial resolution.
Finally we apply an adaptive spatial gaussian filter in the outer
parts: each spectrum is replaced
by the weighted sum of itself and its neighbours with a gaussian width
adapted to the local S/N. At the edges of the field, this process
increases the S/N to $\sim$ 30 - 50.

The reconstructed frame built from the mean intensity of all merged
spectra in the 5300 \Ag - 5400 \Ag wavelength range is presented in
Fig. \ref{fig:recon}.
The spatial resolution of the merged exposure,
derived from the minimization process already described, is
0.87 arcsec. FWHM (Table \ref{tab:seeing}).
The new zero point of the coordinate is set to the precise location
of the nucleus center measured from the C photometric model
(Sect. \ref{par:const} Table \ref{tab:const}) and the offset between
the merged exposure and the HRCAM I frame (Table \ref{tab:seeing}).
Observe that the offcentering of the bright peak with respect to
the photometric center is still present in this reconstructed
image.

\subsubsection{Stellar velocity field}
\label{par:velo}

The stellar velocity field computed from the merged spectra is
presented in Fig.\ref{fig:velo}. Other velocity fields derived
from the individual exposures look
quite similar except for some individual erratic points. In the
central region (two arcsec. radius) there are no significant differences
between the three individual velocity fields and the one presented in
Fig. \ref{fig:velo}. The systemic velocity was taken as that
of the global symmetric point in the velocity field.
%
\begin{figure*}
\fig{fig14.ps}{16}
\caption[]{
Stellar velocity field.
The horizontal axis has been aligned with the nucleus photometric
major axis (PA = 54\degr). Image has normal celestial handedness.
Location of the nucleus photometric center is marked with a white circle.
Zero isovelocity is plotted in dashed line and isovelocity step is 20 \kms.}
\label{fig:velo}
\end{figure*}

For clarity, the velocity field has been filtered
according to the intensities of the reconstructed
frame (Fig. \ref{fig:recon}): the central region with intensities
higher than 30 \% of the maximum intensity have been filtered using
an adaptive scheme (filter/adapt MIDAS command) in a 0.7 arcsec. window
and the rest of the frame was filtered using a 1.1 arcsec. window.
We checked that this process do not alter the velocity field except
in the outer parts were the local signal to noise ratio is smaller.

The observed central velocity field is extremely compact:
the region where V $\ge$ 20 \kms is limited
to 17 pc along the major axis and 7 pc along the minor axis. It
presents a steep
gradient  (45 $\mbox{km.s}^{-1}\mbox{.pc}^{-1}$) in the center and
a sharp drop to almost zero outside the nucleus. The fact that
such large gradients are seen in the experimental data, in spite of the very
effective smoothing effect of both the integration along the line of
sight and the convolution by the PSF, does attest for even more extreme
gradients in the true spatial law (see Sect. \ref{par:svelo} below).
The maximum velocity (125 \kms) is reached at
3 pc from the dynamical center. These results have been
checked for consistency,
using the velocity field computed from the November 1990's
observations, which shares the same features.

Our data, when restricted to the major axis,
are in good agreement with
K88 and DR88 data (Fig.\ref{fig:slit}).
The larger velocity gradient present in our data is simply
due to a better spatial resolution.
%
\begin{figure}
\fig{fig15.ps}{6.0}
\caption[]{
Comparison of major axis stellar velocity profiles according different
authors with the TIGER's data. All spatial elements included in a
slice of 0.5 arcsec. width are presented (full circles).}
\label{fig:slit}
\end{figure}

As noticed by K88 and DR88,
the kinematical center defined as the center of symmetry
of the velocity field, does not coincide with the maximum of light,
but is offcentered with respect to the brightest peak. On the other
hand it is nearly coincident with
the nucleus photometric center (with a small offset of 0.1\arcsec SW
along the nucleus major axis).
We will discuss this point in more detail in Sect. \ref{par:off}.

\subsubsection{Derivation of spatial velocities}
\label{par:svelo}
The observed velocities in the central arcsec. are clearly
dominated by seeing and projection effects. Derivation of the true
spatial velocities is thus a prerequisite to any discussion about the
nature of the nucleus.

The multi-Gaussian expansion technique (Monnet et al., 1992) which has
been successfully applied to the deconvolution and deprojection of the
light in Sect. \ref{par:pproj} and \ref{par:pspac}, can
also be used, with some restrictions, to get the spatial velocity. In
this formalism, the spatial velocities are computed from a sum of
Gaussians functions:
\begin{equation}
\eta_j = C_j \exp{\left(-\frac{\left(x^2+y^2+z^2/q_j^2\right)}
{2 \sigma_j^2}\right)}
\end{equation}
with :
\begin{equation}
\left\{   \begin{array}{l}
\displaystyle{
V_x = - \frac{1}{\nu} \sum_i \sum_j \frac{\partial}{\partial y} \left(
\nu_i \eta_j \right)} \\
\displaystyle{
V_y = \frac{1}{\nu} \sum_i \sum_j \frac{\partial}{\partial x} \left(
\nu_i \eta_j \right) } \\
V_z = 0
\end{array} \right.
\end{equation}
where $\displaystyle{\nu = \sum_i \nu_i}$ is the MGE spatial
intensity computed in Sect. \ref{par:pspac}.

The method is comparable to the one used to deconvolve and deproject
the light intensity. Unfortunately, the inverse problem of the
projection of the spatial velocity cannot generally be solved analytically.
We must then compute the projected and convolved velocities
from the $\eta_j$ Gaussians and compare these values with the
observed data.

As a first approximation, we do not take into account the offcentered
bright peak. In that case, gaussian components corresponding to
the bulge and nucleus
are concentric (see Table \ref{tab:obspace}). Projection and convolution of
velocities is then fully analytical so that the minimization process
between the observed velocities and the models is quite efficient.
We adopted a single gaussian PSF with 0.87 \arcsec FWHM, as
derived in Sect. \ref{par:seeing}.

The spatial parameters of the resulting best fit MGE model
are given in Table \ref{tab:velospace}.
Its predicted major axis velocity profiles, with and without
convolution by the PSF, are shown in Fig. \ref{fig:decvelo}.
Note the very good agreement between the observed and the
computed profiles.
The corresponding spatial velocities reach
340 \kms in the equatorial plane. This value is probably a lower limit;
a nearly equally good fit of the observed velocities has been
obtained with a spatial velocity greater than 900 \kms
in the central 0.1 arcsec. (Fig.\ref{fig:decvelo}).
This fully confirms that the central velocity gradient is not resolved
with our 0.87 arcsec. resolution.
Exploration
of a wide range of models gives an absolute maximum velocity of
150 \kms at r = 2 arcsec.
%
\begin{table}
\begin{flushleft}
\begin{tabular}{rrrr}
\hline
& $C$ & $\sigma$(\arcsec) & q \\
\hline \hline
1 & -1.072 $10^4$ & 1.07 & 1.00 \\
2 & 1.128 $10^4$ & 1.11 & 0.95 \\
3 & -7.588 $10^2$ & 2.75 & 0.46 \\
\hline
\end{tabular}
\end{flushleft}
\caption[]{MGE velocity spatial parameters
(axisymmetric model). $C$ is in \kms.arcsec.}
\label{tab:velospace}
\end{table}
%
\begin{figure}
\fig{fig16.ps}{6.37}
\caption[]{
Major-axis profiles of observed velocities (solid line),
MGE model (short dashed line) and
deconvolved model (long dashed line). Spatial velocities
in the equatorial plane are presented in the inserted panel:
MGE model (solid line) and keplerian model (dashed line).}
\label{fig:decvelo}
\end{figure}

We must then check the perturbation produced by the bright peak on the
projected and convolved velocity field. Using all
light components, including the bright peak, and
the spatial velocities derived previously (Table \ref{tab:velospace})
we integrate numerically the velocities
along the line of sight.
The seeing convolution is also performed numerically.
The perturbation due to projection is larger
($\mbox{V}_{\mbox{perturb.}}
- \mbox{V}_{\mbox{init.}} \simeq +20$ \kms)
when the bright peak is located at the smallest distance of the nucleus
(i.e $z_c = 0$) than when it is
located outside the nucleus
($\mbox{V}_{\mbox{perturb.}}
- \mbox{V}_{\mbox{init.}} \simeq -10$ \kms).
Such an asymmetry is
indeed observed in our velocity field (+15 \kms between the two
sides and 0.1 arcsec. shift) and suggests both that the bright peak
is located inside the nucleus and
that the true kinematical center is closely aligned
with the nucleus geometrical center.

\subsubsection{Stellar velocity dispersion field}

The spectral broadening functions, derived from the FCQ technique, do show
some small oscillating wings, but their cores are very nearly gaussians
at our spectral resolution (101 \kms)
(see Fig. \ref{fig:broad} for a few typical examples).
We checked that the Wiener filter used in the deconvolution process
did not attenuate asymmetries of the broadening functions by replacing it
with a large band low pass filter
(five times the sigma of the cross-correlation function).
As expected, the computed broadening functions present stronger
oscillations in the outer part, but no asymmetry in its core.
%
\begin{figure}
\fig{fig17.ps}{14}
\caption[]{
Broadening functions (solid lines) and gaussian fit (dashed lines)
at three locations: V (0.06, 0.08), P (-0.53, 0.20) and S (0.72, -0.08).
Coordinates are given in arcsec. with the same orientation than in the
previous figures.}
\label{fig:broad}
\end{figure}

The velocity dispersion field computed from the gaussian fit of the
broadening functions is given in Fig. \ref{fig:disp}.
This image has been filtered like the velocity field in
Sect. \ref{par:velo} with the following parameters:
a 0.7 arcsec. window was used for the central region
where intensities of the reconstructed image are
greater than 60\% of the maximum and a 1.1 arcsec. window for the rest of the
field.

The stellar velocity dispersion field presents a
marginally resolved (FWHM $\sim$ 1.4 \arcsec) peak with a maximum
of 230 \kms above a plateau of 150 - 170 \kms. Figure \ref{fig:disp}
reaveals another striking asymmetry:
{\em the dispersion peak is
neither centered on the kinematical center, nor on the photometric
bright peak}, but is located (0.73\arcsec SW)
with respect to the nucleus photometric center. This is illustrated
in Fig.\ref{fig:broad}, where the broadening function near the
velocity dispersion peak (S) is larger than at the kinematical center
(V) or at the brightest light peak (P).

Comparison with K88 and DR88 major-axis data is given in Fig.\ref{fig:disp_g}.
Agreement is good on the overall, but with a larger scatter than in
velocity profile comparison. The value of the maximum velocity dispersion
is comparable to the one measured by K88 and DR88, but its location
agrees better with DR88 than with K88. However, all profiles, including
K88's one, are disymetric with respect to the photometric center and
support, at least qualitatively, the offcentering of velocity dispersion
peak. In the outer parts ($r \geq 2\arcsec$.) our velocity dispersion
is systematically 20 \kms higher: this might be due to our larger
instrumental broadening (100 \kms versus 25 \kms in DR88
and 58 \kms in K88) which classically leads to overestimate
the lower dispersion values.

%
\begin{figure*}
\fig{fig18.ps}{16}
\caption[]{
Stellar velocity dispersion field.
The horizontal axis has been aligned with the nucleus photometric
major axis (PA = 54\degr). Image has normal celestial handedness.
Location of the nucleus photometric center is marked with a white circle.
Maxium isovelocity dispersion and isovelocity dispersion step are
respectively 220 and 10 \kms.}
\label{fig:disp}
\end{figure*}
%
\begin{figure}
\fig{fig19.ps}{5.0}
\caption[]{
Comparison of major axis stellar velocity dispersion
profiles according to different authors with the TIGER's data (unfiltered).
All spatial elements included in a
slice of 0.5 arcsec. width are presented (full circles).}
\label{fig:disp_g}
\end{figure}

\subsubsection{Non-centered second-order moments}
\label{par:mu2}
The virial theorem will be used later in Sect. \ref{par:vir} to derive
the mean mass-to-light ratio of the nucleus,
independently of specific kinematic models. We will then need
the projected radial non-centered
second-order moment ($\overline{\mu^2}$) which is by definition:
\begin{equation}
\conv{\mu^2} = \conv{\sigma^2} + \conv{V}^2
\end{equation}
$\conv{V}$ and $\conv{\sigma}$ being respectively the observed
velocity and velocity dispersion.
The mean value of this parameter defined as:
\begin{equation}
\label{eq:defmu2}
\ave{\mu^2} = \frac{\dint_{-\infty}^{\infty} \conv{\nu} \conv{\mu^2}
\diff \proj{x} \diff \proj{y}}{\dint_{-\infty}^{\infty}
\conv{\nu} \diff \proj{x} \diff \proj{y}}
\end{equation}
is related to the kinetic energy and will be used to compute
the mean M/L through the virial theorem.

In principle we would have to integrate $ \conv{\mu^2} $
over the whole galaxy, but as we are interested only in the M/L of the
nucleus, we can limit the integration to the nucleus itself.
Integrating Eq. \ref{eq:defmu2} inside
an ellipse with a major-axis and minor-axis radii of respectively 3\arcsec
and 2\arcsec,
centered and aligned on the nucleus, we get a {\em local} mean second
order radial moment
$ \ave{\mu^2_{\mbox{\small{T}}}} =
\left(185\right)^2 \mbox{km}^2 \mbox{.s}^{-2} $ (the subscript T
indicating that this value is the sum of the nucleus and bulge
contributions).

If we assume that the nucleus is dynamically isolated from the bulge, one can
go
further and compute $\ave{\mu^2_{\mbox{\small{N}}}}$. In that case,
we have:
\begin{equation}
\conv{\nu}_{\mbox{\small{N}}} \conv{\mu^2}_{\mbox{\small{N}}} +
\conv{\nu}_{\mbox{\small{B}}} \conv{\mu^2}_{\mbox{\small{B}}} =
\conv{\nu}_{\mbox{\small{T}}} \conv{\mu^2}_{\mbox{\small{T}}}
\end{equation}
so that:
\begin{equation}
\ave{\mu^2_{\mbox{\small{N}}}} =
\frac{\dint_{-\infty}^{\infty} \left(
\conv{\nu}_{\mbox{\small{T}}} \conv{\mu^2}_{\mbox{\small{T}}} -
\conv{\nu}_{\mbox{\small{B}}} \conv{\mu^2}_{\mbox{\small{B}}} \right)
\diff \proj{x} \diff \proj{y}}{\dint_{-\infty}^{\infty}
\conv{\nu}_{\mbox{\small{N}}} \diff \proj{x} \diff \proj{y}}
\end{equation}
The integrand is null outside the nucleus because
$\conv{\nu}_{\mbox{\small{T}}} \conv{\mu^2}_{\mbox{\small{T}}} \sim
\conv{\nu}_{\mbox{\small{B}}} \conv{\mu^2}_{\mbox{\small{B}}}$ in
that region. Furthermore, in the nuclear region, the
bulge velocity contribution to $\conv{\mu^2}_{\mbox{\small{B}}}$
is negligible and the bulge velocity dispersion is approximatively
constant ($\conv{\mu^2}_{\mbox{\small{B}}} \sim
\left(150\right)^2 \mbox{km}^2 \mbox{.s}^{-2}$). We finally get:
\begin{eqnarray}
\ave{\mu^2_{\mbox{\small{N}}}} & = &
\frac{\dint_{\mbox{\small{N}}}
\conv{\nu}_{\mbox{\small{T}}} \conv{\mu^2}_{\mbox{\small{T}}}
\diff x \diff y - \left(150\right)^2 \dint_{\mbox{\small{N}}}
\conv{\nu}_{\mbox{\small{B}}}
\diff x \diff y}{\dint_{\mbox{\small{N}}}
\conv{\nu}_{\mbox{\small{N}}} \diff x \diff y} \nonumber \\
& = & \left(250\right)^2 \mbox{km}^2 \mbox{.s}^{-2}
\end{eqnarray}

It should be observed that this quantity which is of great
importance for the following discussion, cannot
be derived from single velocity and velocity dispersion
profiles. This makes it difficult to derive masses through
the virial theorem from
classical long-slit kinematical observations.

\subsubsection{Ionized gas}

The two "red" exposures have been processed exactly as the "blue" ones,
except that they were flux calibrated with an exposure on HD 19445.
Spectra were continuum substracted to search for possible emission
lines.

Apart from the strong \Halpha~ absorption, no feature are observed
in these wavelength range. In particular we do
not detect \Halpha~ or [NII]
emission in the 8 x 6.5 $\mbox{arcsec}^2$. field (normal celestial
orientation). A simple simulation shows that we can put an
upper limit of 7.\ten{-23} \Wm on the central flux in the [NII] 6583.6~\Ag
line. \Halpha~ emission is much more difficult to detect because of
the very deep \Halpha~ absorption.

\section{Discussion}
\subsection{Shape and mass distribution of the nucleus}

In that section, and as a first step towards the investigation
of the nucleus, we will neglect the various
offsets observed in the nucleus. Their implication will be
discussed in the next section.

\subsubsection{A stellar bar ?}

Gerhard (1986) has explained the observed 25\degr~ twist
between the nucleus and the disk minor-axis
by invoking a strongly triaxial nuclear bar sharing the same
equatorial plane than the disk. Gerhard (1988) has then suggested that
a set of adequate orbits could eventually build a tumbling bar model,
compatible with the observed velocity dispersion and rotation velocity
on the major axis, and with an \ML{V} ratio of 8 only
(corresponding to \ML{I} $\sim 1.4$).

Interestingly enough, our two-dimensional stellar
velocity field offers a sharp test of such a model. With an apparent
twist $\psi$ = 25\degr~ between the photometric
minor axis of the nucleus and the disk, the kinematical minor
axis of the nucleus would then be tilted w.r.t. its photometric minor
axis by an angle $\gamma$ given by (Monnet et al. 1992, Eq. 41):
$\tan \gamma = - \tan \psi / \conv{q}^2$, where $\conv{q}$ is the observed
photometric axis ratio of the nucleus.
Using the average axis ratio of the nucleus ($\conv{q}$ = 0.61) gives
$\gamma = - 27\degr$, while
its upper limit from Fig. \ref{fig:velo} is $\sim \pm 5\degr$.
The expected corresponding velocity field, according to the MGE
photometric model (Table \ref{tab:const}) is presented in
Fig. \ref{fig:bar} and is definitively not compatible with our
observations.
Such a low mass tumbling bar model can therefore be discarded.
\begin{figure}
\fig{fig20.ps}{6.00}
\caption[]{
Iso-velocities of the bar model. Note the tilt between the
zero isovelocity (dashed lines) and the nucleus photometric minor-axis
(Y axis). Isovelocity step is 25 \kms.}
\label{fig:bar}
\end{figure}

\subsubsection{Mass to light ratio of the nucleus
from the virial theorem}
\label{par:vir}
The tensor virial theorem offers a direct way to compute the mass to
light ratio of the nucleus, independently of the detailed
kinematics of the stars, provided that it can be considered as
an isolated stable entity (i.e. satisfies Liouville equation).
Given the abrupt decrease of the spatial light intensity
and of the mean rotation velocity at the edge of the nucleus,
this seems a very reasonable assumption.
This point is strongly enhanced by the fact that the maximum
correction factor, derived from a local version of the tensor
virial theorem (see Appendix \ref{par:app1}), is of the order of
1\%, hence fully negligible.

Emsellem et al. (1993a) have shown that for a triaxial galaxy
with cylindrical rotation and whose light
distribution is given by a sum of Gaussians with constant axial ratios
$t$ and $u$, M/L is given by:
\begin{equation}
\label{eq:ml}
M/L = 74 \frac{\ave{\mu^2}}{t u N}
\frac{\displaystyle{ \sum_j \sigma_j^3 I_j}}
{\displaystyle{ \sum_i \sum_j \sigma_i^3 \sigma_j^3 I_j I_i
\left(\sigma_i^2 + \sigma_j^2\right)^{-1/2}}}
\end{equation}
where the M/L ratio in $\mbox{M}_{\sun} \mbox{/} \mbox{L}_{\sun}$ is
supposed to be constant over the nucleus and $\ave{\mu^2}$ in
$\mbox{km}^2\mbox{.s}^{-2}$ is the non-centered second-order radial
velocity moment averaged over the nucleus.
The light distribution is represented by a sum of Gaussians, with spatial
radii $\sigma_j$ in pc and spatial intensities $I_j$ in
$\mbox{L}_{\sun}\mbox{pc}^{-3}$ and with axial ratio $t, u$ (identical
for all Gaussians). $N$ is an adimensional geometrical parameter, given by:
\begin{equation}
N = \frac{A_1}{t^2} \sin^2\theta \sin^2\phi +
\frac{A_2}{u^2} \sin^2\theta \cos^2\phi +
A_3 \cos^2\theta
\end{equation}
where $\theta$ and $\phi$ are the first two Euler angles of the line of
sight, with respect to the principal axis $\mbox{O}_{x,y,z}$ of a triaxial
galaxy rotating around $\mbox{O}_z$.
The $A_i$s are functions of t and u only and can be found in the
Table 2.1 p. 57 of Binney and Tremaine (1987).

We first compute the M/L ratio in the axisymmetric (oblate) case, using
the spatial model E, given
in Table \ref{tab:ellspace}. An inclination $\theta$ of 77\degr~ has been
assumed.
$I$ and $\sigma$ are converted respectively in $\mbox{L}_{\sun}\mbox{pc}^{-3}$
and in pc, assuming a distance of 700 kpc.
Classically, with the usual notation $e = \sqrt{1 - q^2}$
($q < 1$) we have:
\begin{equation}
A_1 = A_2 = \frac{q}{e^2} \left(\frac{\arcsin e}{e} - q\right) = 0.516
\end{equation}
\begin{equation}
A_3 = \frac{2 q}{e^2} \left(\frac{1}{q} - \frac{\arcsin e}{e}\right) =
0.968
\end{equation}
and:
\begin{equation}
t u N = A_1 \sin^2\theta + q^2 A_3 \cos^2\theta = 0.507
\end{equation}
$\ave{\mu^2}$ has been computed in Sect. \ref{par:mu2}
from the observational data. It is equal to
$\left(250\right)^2 \mbox{km}^2\mbox{.s}^{-2}$, which gives
\ML{I} = 69.4.
For an exact edge-on system, the M/L ratio would only be
slightly changed to \ML{I} = 68.2.

We have seen in the previous section, that the axisymmetric
nature of the mean rotation velocity field restricts acceptable triaxial
models to the ones with one axis closely parallel to the line of sight.
Such models are rather contrived, relying on a fortuitous
alignement, with an a priori probability of $\sim 5\%$ only,
but nevertheless cannot be excluded completely.

To bracket the possible M/L ratios, we consider two extreme cases:
\begin{itemize}
\item a prolate nucleus, with one of its short axis parallel to the
line of sight ($t=1$, $u=\proj{q}$)
\item an highly elongated nucleus with its longest axis, parallel to the
line of sight with $t=0.25$, $u=\proj{q}$
\end{itemize}
For simplicity, we consider exact edge-on models only.
The parameters used in Eq. \ref{eq:ml} are then
$\theta=\frac{\pi}{2}$, $\phi=\frac{\pi}{2}$, and $tuN=A_1\frac{u}{t}$.
In the prolate case:
\begin{equation}
A_1 \frac{u}{t} = \frac{\conv{q}^3}{\conv{e}^2} \left[
\frac{1}{\conv{q}^2} - \frac{1}{2 \conv{e}} \log \left(
\frac{1+\conv{e}}{1-\conv{e}} \right) \right] = 0.480
\end{equation}
and \ML{I}= 73.3.

In the elongated case:
\begin{equation}
A_1 \frac{u}{t} = 2 t \frac{\mbox{F}\left(\xi,k\right) -
\mbox{E}\left(\xi,k\right)}{k^2 \sin^3\xi}
\end{equation}
with $\xi = \arccos t = 75.52\degr$ and
$k = \frac{1}{u} \sqrt{\frac{u^2 - t^2}{1 - t^2}} = 0.942$, and
where F and E are respectively the incomplete elliptic integrals of first
and second kind. This gives $A_1 \frac{u}{t} = 0.479$ and
\ML{I} = 73.5.

One sees that there is very little
dependence between the geometrical shape, and the derived
M/L ratio. We can thus be reasonably sure that there
is indeed a large jump in mass to luminosity ratio in the
center of M31, at least within our present set of hypothesis, i.e.
for arbitrary non-tumbling triaxial models. In Appendix \ref{par:app2},
the case of a tumbling bar is quantitatively
studied, and a firm limit of -16\%  on the corresponding
correction to the derived M/L ratio is found. This seems to
settle the last possible loophole and establishes a high mass-to-light
ratio in the nucleus of M 31.

\subsubsection{Dust obscuration ?}

This very high \ML{I} jump from $\sim$ 2.6 in the bulge
to $\sim$ 70 in the nucleus could in principle be due either to a mass
increase in the nucleus (the most common hypothesis since this
phenomenon has been observed in the early 70s), or conversely to a
large decrease of the observed luminosity, because of strong dust absorption
in the nucleus. Quantitatively, with a total luminosity of the nucleus
of 6.5 \ten{6} \Lsun, this would require a total absorption of
1.6 \ten{8} \Lsun, which would be reradiated in the infrared,
probably around 100 \micron.

Soifer et al (1986), using the IRAS satellite, have indeed observed an
infrared excess of 7.1 \Jansky at 60 \micron and 12.0 \Jansky at 100 \micron
in the central part of M31, with beams approximately 2 arcmin. across or more.
This gives a colour temperature of 45~\Kelvin and a total radiated flux,
assuming a pure blackbody type spectral emission of 6.9 \ten{6} \Lsun. Their
measures in a number of apertures centered on the galaxy, show a general
slope close to that of the bulge star's light, so that even this
small flux, compared to the required reradiated flux of 1.6 \ten{8} \Lsun,
must mainly originate in the bulge itself, and only a small fraction
of it can come from the nucleus. We can thus be reasonably sure that
there is indeed a genuine excess of mass in the center of M31.

\subsubsection{A massive nucleus ?}

The simplest interpretation is then that the stellar
population in the nucleus has a much higher global mass-to-luminosity
ratio than the bulge, presumably because of atypic stellar formation
connected with its unique physical environment. However, as shown by
Mould et al. (1989), the remarkable lack of any color excess in the
nucleus, and especially in V-K, precludes low-mass stars exceeding the
main-sequence hydrogen burning limit as a viable explanation. In that
frame, we are thus forced to  a rather contrived  hypothesis, namely
a strong excess of brown dwarfs and/or planetary bodies, without any
significant enhancement in the proportion of more conventional M stars.

We have then run self-gravitating isotropic models
using the
axisymmetric nucleus and bulge light density model (model A, Table
\ref{tab:axispace} in Sect. \ref{par:pspac}) with bulge \ML{I} = 2.56
and nucleus \ML{I} ranging from 2.56 to 60.
The technique used to build such
models is detailed in Emsellem et al. (1993a)
and briefly described in Appendix \ref{par:dynmod}.
Projected velocities
and velocity dispersion were convolved with a 0.87\arcsec Gaussian
according to the measured spatial resolution of TIGER's data (Sect.
\ref{par:seeing}). Major-axis velocity and velocity dispersion
profiles are presented in Fig. \ref{fig:isotrop}
\footnote{For reason of clarity we present only the extracted
major-axis profiles of velocity and velocity dispersion fields}.

The experimental data exhibit three main features, a strong velocity
dispersion peak and a strong velocity gradient in the inner one arcsec., as
well as a sharp drop of the rotation velocity outside the nucleus. Clearly,
none of the models can represent more than one of these characteristics.
Anisotropic models
can, of course, be called upon: with a central radial anisotropy,
we can potentially match the central velocity dispersion peak, but
the predicted rotation velocity gradient would be even smaller than
the one shown in Fig. \ref{fig:isotrop}.
With a central tangential anisotropy, the
observed central rotation velocity gradient could certainly be matched,
but now at the expense of a much decreased velocity dispersion towards
the center. It is thus clear that the mass excess must necessarily
have a core radius significantly smaller than that of the nucleus
itself. While not an ironclad proof, this certainly points to
a supermassive black hole (SBH) as, by far, the most likely explanation.
\begin{figure}
\fig{fig21.ps}{10}
\caption[]{
Self-gravitating isotropic models with constant M/L.
Bulge \ML{I} is set to 2.56, while nucleus \ML{I} varies from 2.56 to
60.
Upper panel: Major-axis velocity
profile. Lower panel: Major-axis velocity dispersion profile. Filled
symbols: TIGER data. Open symbols: K88 data.}
\label{fig:isotrop}
\end{figure}

\subsubsection{A supermassive black hole ?}
We have then computed an isotropic oblate model
for identical M/L ratios in the bulge and the nucleus (\ML{I} = 2.56)
and an SBH mass of 1.8 \ten{8} \Msun:
the rotation curve is well matched,
but the predicted velocity dispersion curve is much too high
(Fig. \ref{fig:sbh}).
\begin{figure}
\fig{fig22.ps}{10}
\caption[]{
Self-gravitating dynamical models including a supermassive black hole.
Bulge and nucleus \ML{I} are equal to 2.56. SBH mass and
shape of the velocity ellipsoid are indicated in the figure.
Upper panel: Major-axis velocity profile.
Lower panel: Major-axis velocity dispersion profile.
Filled symbols: TIGER data. Open symbols: K88 data.}
\label{fig:sbh}
\end{figure}

An improvement of the modelling
requires some degree of tangential anisotropy
(i.e. to set the principal axis of the velocity ellipsoid $\sigma_\theta^2$
in the tangential direction smaller than the $\sigma_z^2$ axis in the azimuthal
direction). In this case, kinetic energy is transferred from the
random ($\sigma_{\theta}$) to the ordered motions ($V$).
To compute the maximum effect of such an anisotropy, we set
$\sigma_{\theta} = 0$ and $\sigma_z = \sigma_r$ in the nucleus while
the bulge remains isotropic. The corresponding anisotropic profile is
shown in Fig. \ref{fig:aniso} (see also Appendix \ref{par:dynmod})
and velocity and velocity
dispersion profiles in Fig. \ref{fig:sbh}.
Note that this model match well the observed velocity
while the velocity dispersion peak is narrower than the observed one.
It required, however, a very strong and probably
unphysical anisotropy in the central part
with $\sigma_{\theta}^2 \simeq 0$.
Because our model is restricted to the assumed and somewhat
arbitrary cylindrical geometry of the
velocity ellipsoid, we do not attempt
to explore all the parameter space (mass-to-light
ratios, SBH mass and anisotropic
parameters).
This model gives, however, a
better estimate of the black hole mass, at 7.2~\ten{7}~\Msun~ and
proves the need for a strong tangential anisotropy in the nucleus.

\begin{figure}
\fig{fig23.ps}{6}
\caption[]{Shape of the velocity ellipsoid corresponding to the
anisotropic dynamical model. This profile display the value
of the anisotropic parameter ($\alpha = 0, \beta = 4$)
$\sigma_{\theta}^2$/$\sigma_z^2$ in the equatorial
plane (see appendix \ref{par:dynmod} for explanation)}
\label{fig:aniso}
\end{figure}

An alternative to the SBH hypothesis is a dense stellar cluster
lying at the center of the nucleus. To test this suggestion
we ran a set of models with various central mass concentrations
by representing the stellar cluster density distribution
as a Gaussian of varying width $\sigma_c$.
Comparison with the experimental data gives
$\sigma_c \la 0.7 \mbox{pc}$.
Note that Richstone et al. (1990) developed a similar
analysis of the DR88 data and found a corresponding maximum
width of 1.15 pc.
This larger value can be explained partly by our
slightly better spatial resolution
\footnote{
True spatial resolution of DR88 is unknown, but it can not be
smaller than their slit width (1 arcsec.).}
as well as their oversimplifying hypothesis of spherical symmetry.

The anisotropic model with a high central mass concentration gives
finally the
best fit to the observed stellar velocity field (Fig. \ref{fig:velo2d}).
However, it is not able to
reproduce the offcentered light peak, nor the observed offcentered
velocity dispersion field (Fig. \ref{fig:sig2d}).
These strong offsets
could well invalidate the previous ``naive''
axisymetric analysis and are investigated in the next section.
\begin{figure}
\fig{fig24.ps}{8}
\caption[]{Stellar velocity field of the anisotropic model
with a 7.2 \ten{7} \Msun supermassive black hole (isovelocities)
compared to the observed one (grey level). Isovelocity step is 20 \kms.
}
\label{fig:velo2d}
\end{figure}
\begin{figure}
\fig{fig25.ps}{8}
\caption[]{Stellar velocity dispersion field of the anisotropic model
with a 7.2 \ten{7} \Msun supermassive black hole (isovelocities)
compared to the observed one (grey level). Isovelocity step is 20 \kms.
}
\label{fig:sig2d}
\end{figure}

%
\subsection{Asymetries in the nucleus}
\label{par:off}

The outer isophotes of the nucleus have a well-defined center of
symmetry, marked N on Fig. \ref{fig:off}. However, as discovered by
Light et al. (1974) from a V band image taken during a Stratoscope II
flight, the sharp central light peak P is displaced by 1.4 pc. N-E,
roughly along the major axis of the nucleus (see Fig. \ref{fig:off}).
Since then, combined results from, e.g. Nieto et al. (1986),
Crane and Stiavelli (1993) and Mould
et al. (1989), have shown that this photometric asymmetry is present
in a large wavelength range, from the UV (0.18\micron) to
the near-IR (2.2\micron).

High spatial resolution radial velocity profiles along the major-axis,
obtained by K88 and DR88, have
shown that the kinematical center of symmetry V is not located at the
light peak, but instead at, or very close to, the photometric center
of symmetry N (Fig. \ref{fig:off}). Our two-dimensional
radial velocity field map (Fig. \ref{fig:velo}) fully confirms this results:
it appears quite symmetric, centered with respect to N, and does not
show any significant disturbance at P except the slight asymmetry
already discussed in Sect. \ref{par:svelo}.
We have also obtained a two dimensional map of the radial velocity
dispersion field (Fig.~\ref{fig:disp}). It exhibits a well-defined
central peak S, located roughly symmetrically
of P with respect to N (Fig. \ref{fig:off}).

We see that, observationally, the center of M31 harbors a number of
puzzling asymmetries. We explore possible explanations in the
next section.
%
\begin{figure}
\fig{fig26.ps}{4.85}
\caption[]{
Schematic view of the location of the nucleus components.
N, $\mbox{R}_{\mbox{N}}$ and $\mbox{M}_{\mbox{N}}$ are respectively
the photometric nucleus center, the deconvolved core isophote and
the major-axis.
$\mbox{M}_{\mbox{B}}$ is the photometric major-axis of the bulge.
P, $\mbox{R}_{\mbox{P}}$ are respectively the photometric bright peak
center and deconvolved core isophote.
V is the stellar rotation center and S the location of the maximum
stellar velocity dispersion. Orientation is North (up) and East (left).}
\label{fig:off}
\end{figure}

\subsubsection{Dust effect ?}

Obscuration by dust filaments, non uniformly distributed near
the nucleus, could possibly reproduce the light asymmetry, and
was indeed first proposed by Light et al. (1974). The remarkable
constancy of this asymmetry, from UV to IR, and especially
in Mould et al. (1989) 2.2\micron observations, makes that
hypothesis untenable.

Another explanation proposed by Mould et al. (1989), would be a disk
or torus of dust embedded within the nucleus and optically thick,
even at 2.2\micron. Such a structure, tilted along the line of sight
(by 30\degr~ in the Mould et al. simple model) could potentially
reproduce the photometric appearance.

The nucleus shows a highly contrasted velocity dispersion peak
at S. If this dynamical appearance was due to a dust effect only,
this would necessarly be also the region closest to the center,
i.e. it would be at the location of the light peak.
Just the opposite is in fact observed, so that
this model is inconsistent with the
kinematical data.

\subsubsection{A gravitational lens ?}

Since the central bright peak does not exhibit any spectral peculiarity
w.r.t. the nucleus itself, it could a priori be due to a gravitational
amplification of the nucleus by a compact object located between us and
M31. Indeed, using the formalism of Bontz (1979), one easily finds
that an halo
object, located at 10 kpc and with a mass of 5.~\ten{4} solar masses, would
produce the observed amplification by of about a factor 1.4.

However, this amplification then arises solely inside the Einstein radius of
the deflector (of about 0.05 arcsec. in that case), and is exactly compensated
in total flux by a decrease in surface brightness outside it. A gravitational
amplification is therefore
not able to reproduce the observed single peak, and in
fact would classically give a partial ring.

\subsubsection{A stellar cluster ?}

As already discussed by DR88 and Mould et al. (1989),
this offset bright peak can correspond to an individual stellar clump,
a globular cluster or the remains of a recently captured dwarf galaxy
orbiting around the nucleus. Note that the absence of color gradients
and emission lines precludes significant stellar formation
to occur in this clump.

On the other hand, its physical properties well match those of a globular
cluster: from Harris and Racine (1979), the absolute B magnitudes of M31
globular clusters are $\mbox{M}_{\mbox{B}} = -7.3 \pm 1.2$, which
translates into $\mbox{M}_{\mbox{I}} = -10.4 \pm 1.2$, in good agreement
with the absolute magnitude of the peak
($\mbox{M}_{\mbox{I}} = -10.7$, Table \ref{tab:global}). Furthermore,
according to Peterson and King (1975), their mean core radius is
$1.3 \pm 0.6$ pc, similar to the upper limit (1.2 pc) obtained on the
bright peak.

This explanation faces, of course, a rather severe problem, as already
pointed out by DR88 and Mould et al. (1989), namely
that the lifetime of the system, if orbiting at 1.5 pc from the center, is
only $\sim$ \ten{7} years only. Alternatively, the lopsided photometric
shape of the nucleus could be due to a chance projection effect, with the
globular cluster orbiting safely well outside the nucleus. However, the
lifetime of such a configuration would typically be $\sim$ \ten{6} years,
and its a priori probability is $\sim$ \ten{-4}.

Even if this explanation appears quite unlikely on statistical grounds,
it would be nice to get a direct test from observational data. This is in
principle tractable using our spectral data: a globular cluster, with an
internal velocity dispersion of $\sim$ 7 \kms, and to a lesser extent
the nucleus of a dwarf galaxy with $\sim$ 70 \kms, will necessarily
modify the shape of the spectral lines.

To test whether this effect is measurable, we proceeded as follows:
\begin{itemize}
\item A composite broadening function is computed as the sum of the
nucleus and the stellar cluster broadening functions, assumed
both to be Gaussians.
The total intensity {$\cal A$} of the Gaussian corresponding to the
stellar cluster is fixed to the observed ratio (0.18) of the peak to the
underlying nucleus intensity measured in the TIGER reconstructed image.
Nucleus
mean velocity and velocity dispersion (respectively 117 and 180 \kms)
are taken from the observed values in the vicinity of the bright peak.
\item The template spectrum previously used for M31 is convolved with
this broadening function
\item Finally the FCQ program is run
on this simulated spectrum using the same template.
\end{itemize}
Figure \ref{fig:glob} and Table \ref{tab:glob} show the results
of this simulation in three cases: a globular cluster located
in the bulge (A) or within the nucleus (B), a nucleus of a dwarf
galaxy in the vicinity of M31's nucleus (C).
As illustrated in panel A and B of Fig. \ref{fig:glob},
it is formally not possible to retrieve a dispersion lower than
the spectral resolution.
However, the output broadening functions
is quite asymmetric at our resolution ($\sigma_{\mbox{instr}}$ = 105 \kms)
when the radial velocity of the globular cluster is
significantly different from the nucleus one (case A). Such an
asymmetry is not present in the data (see the observed broadening
function labelled P in Fig. \ref{fig:broad}), so that this case can
be ruled out.

On the contrary, recovered broadening functions are nearly
gaussian when the object has the same radial velocity than
the nucleus (B and C case), the sole measurable effect being
the lowered output velocity dispersion.
Indeed, the velocity dispersion is lower at P (180 \kms)
than at the symmetric point with respect to N (220 \kms). This would
favor the assumption of a stellar cluster orbiting into M31's nucleus.

%
\begin{table}
\begin{flushleft}
\begin{tabular}{l|llllll|ll}
\hline
Model & \multicolumn{6}{|c|}{Input} & \multicolumn{2}{c}{Output}\\
&  \multicolumn{3}{|c}{Nucleus} & \multicolumn{3}{c|}{Object} \\
& $\cal A$ & $V$ & $\sigma$ & $\cal A$ & $V$ & $\sigma$ & $V$ & $\sigma$ \\
\hline \hline
A & 1 & 117 & 180 & 0.18 & 0 & 7 & 74 & 164 \\
B & 1 & 117 & 180 & 0.18 & 117 & 7 & 117 & 126 \\
C & 1 & 117 & 180 & 0.18 & 117 & 80 & 117 & 153 \\
\hline
\end{tabular}
\end{flushleft}
\caption[]{Simulation of the perturbation of the broadening functions.
$\cal A$ is the Gaussian total intensity, $V$ and $\sigma$ are in \kms}
\label{tab:glob}
\end{table}
%
\begin{figure}
\fig{fig27.ps}{12}
\caption[]{Simulation of the perturbation of the broadening functions.
See text and Table \ref{tab:glob} for details.
Solid lines: input broadening function. Long dashed lines: recovered
broadening functions. Short dashed lines: Gaussian fit of the recovered
broadening functions.}
\label{fig:glob}
\end{figure}

We also proceed to another modelling to check if the perturbation
due to this stellar cluster on the stellar velocity dispersion field
could be responsible of the observed offsets.
We modelled the unperturbed central velocity dispersion field
$\sigma_{\mbox{init}}\left(x,y\right)$ as
the sum of a centered Gaussian of 2.1\arcsec~ FWHM and 100 \kms
maximum value, plus a constant value of 160 \kms.
Using the deconvolved MGE velocity model
($V$, Table \ref{tab:velospace}) as well as the
C light model of the offcentered bright peak ($\nu_{\mbox{p}}$) and the
centered nucleus and bulge components ($\nu_{\mbox{c}}$), we
computed the perturbed velocity dispersion field $\sigma_{\mbox{per}}$
as:
\begin{equation}
\sigma_{\mbox{per}} = \sqrt{\conv{\mu^2}_{\mbox{per}} -
\conv{V}^2_{\mbox{per}}}
\end{equation}
where $\conv{\mu^2}_{\mbox{per}}$ and $\conv{V}_{\mbox{per}}$
are respectively the convolved
non-centered second-order momentum and convolved velocity. These
are computed as:
\begin{equation}
\conv{V}_{\mbox{per}} = \frac{\left(\nu_{\mbox{p}}+\nu_{\mbox{c}}\right)
V \otimes PSF}
{\left(\nu_{\mbox{p}}+\nu_{\mbox{c}}\right)\otimes PSF}
\end{equation}
and
\begin{equation}
\conv{\mu^2}_{\mbox{per}} = \frac{
\left[\nu_{\mbox{p}} \left(V^2 + \sigma_{\mbox{p}}^2\right)
+ \nu_{\mbox{c}} \left(V^2 + \sigma_{\mbox{init}}^2\right)\right] \otimes PSF}
{\left(\nu_{\mbox{p}}+\nu_{\mbox{c}}\right)\otimes PSF}
\end{equation}
The PSF was assumed to be a Gaussian of 0.87 arcsec. FWHM
(see Table \ref{tab:seeing}).
Fig. \ref{fig:glob2}
displays the result of this simulation in two cases $\sigma_{\mbox{p}}$~=~7
and 80~\kms. It shows that
the peak dispersion is only
slightly displaced ($\sim$~0.1~arcsec)
towards the opposite direction of P.
This is a far too small value to explain the observed shift of 0.7 arcsec.
\begin{figure}
\fig{fig28.ps}{8}
\caption[]{Simulated perturbed velocity dispersion profiles
(nucleus major-axis) compared to
the convolved unperturbed initial profile (solid line).
Filled symbols: TIGER data. }
\label{fig:glob2}
\end{figure}

\subsubsection{Dynamical oscillation}

In the preceding paragraphs we checked different hypothesis concerning
the offcentering of the brightest peak relatively to the dynamical center.
All these hypothesis were based on the same scheme: the normal symmetric
and centered nucleus is affected in some way (light obscuration,
light magnification or captured object) to resemble the observed picture.
At the opposite the simplest interpretation of the observational data,
i.e. the offcentering of the nucleus with respect to the center of mass,
is difficult to explain using the classical models.

Comprehensive kinematical modelling, i.e. for a triaxial flattened
non isotropic nucleus, but neglecting the decentering of the various
components,
gives a clear indication of a central dark mass concentration, presumably
a black hole.
However, the dynamics of the central three parsecs is dominated
by the clear decentering of the three detected components,
the luminosity peak P,
the rotation center V and the dispersion center S.

According to Miller and Smith (1992), there is indeed a natural instability
in the nuclear region of galaxies, in the form of an oscillating wave.
This could explain the shift of the maximum of light from the center of
gravity of the galaxy, as well as from its center of rotation.
Miller and Smith (1992) found that this could involve substantial mass
and energy transfer.
This raises the intriguing possibility, that
the observed displacement of the maximum velocity dispersion
and the maximum of light
could be explained in such a framework, perhaps without the need
for a strong mass concentration.
Further discussion along this line is outside the scope of this
paper: we are presently undertaking orbital analysis and full N-body
simulations
to investigate this hypothesis.

\section{Conclusions}
Using the new and high quality observational material provided
by HRCAM and the TIGER spectrograph at CFHT,
we have investigated in detail
the photometric and kinematical properties of the nucleus of M31.
Neglecting, in a first step, the various kinematical and photometric
offsets, we claim that the nucleus cannot be a nuclear rotating
bar with ordinary M/L. On the other hand,
we have obtained strong evidences for a high central mass
concentration in the nucleus, possibly a supermassive
black hole, of about 7.\ten{7} \Msun, as well as a high degree of
tangential anisotropy.
Note that this is
based on self consistent kinematical models which takes
into account the real light distribution, including
flattening, and that, within the above hypotheses, this
high total nuclear mass is confirmed through a comprehensive
virial tensor theorem approach.

The real galaxy, however, does exhibit these puzzling asymmetries,
with a clear peak of light and a clear velocity dispersion
peak, both significantly offsett from the kinematical
center, which also coincides with the geometrical center of the
nucleus. We cannot completely exclude that some part of this
pattern is in fact due to light absorption by a central disk
of dust or a separate stellar system orbiting the nucleus,
but this cannot realistically explain the overall data,
and expecially the offset between the light and the velocity
dispersion peaks. The asymmetric kinematic behaviour of the
nucleus of M31 is thus basically real.
The huge central dark mass concentration (7.2 \ten{7} \Msun)
required from the axisymmetric approach would dominate the
overall mass of the nucleus whose visible part contributes
only for a few \ten{6} \Msun. In that case,
even crude kinematical models
immediately show that no significant offsets are possible, as
the overall dynamics of the nucleus is then totally dominated
by the central black hole.

Lauer et al. (1993) have suggested that a {\em second} supermassive
black hole could be associated with the observed light peak
(a remain of a cannibalized dwarf galaxy nucleus). This should
increase the lifetime of this offset cluster, which thus could
have survived to the tidal forces induced by M31's nucleus.
However, in the light of our bidimensionnal spectrographic
data, it is almost impossible to reconcile such an hypothesis
with both the rather regular stellar velocity field and the
dispersion peak (offset in the {\em opposite} direction
of the presumed massive stellar cluster).

We are thus left to argue that a more direct explanation would
be a natural global oscillation of the stellar component of
the nucleus, \`a la Miller and Smith, as the source of the
light peak offset. The internal dynamics of such a system is
rather complex, but leads quite naturally to the
offset between the kinematical and photometric centers.
Possibly, it could also explain the velocity dispersion peak
as the (transient) unstable lagrangian point of the global
system. Again, this is plausible only if there is no dominant
central mass concentration, which would sweep out any such
effect. Fortunately, this possibility remains open, as
obviously the virial approach used in this paper fails for
such a non stationnary, non axisymmetric situation.

Clearly the dynamical status of the nucleus remains much
uncertain. Sorely needed are fully N-body simulations
to explore its possible state, as well as bidimensional,
0.1-0.2 arcsec. spatial resolution, spectrographic data,
which should be available relatively soon, when adaptive
optics capabilities on large telescopes are in operation.

\acknowledgements{We are grateful to Ralph Bender and Ortwin Gerhard
for helpful discussion. The FCQ C code was written according to the
original Fortran code of Ralf Bender.}

\appendix
\section{Appendix: virial theorem correction in the
case of a non isolated nucleus}
\label{par:app1}
The classical virial method fails if the nucleus is not truly
isolated, but only a local density enhancement, resulting from a {\em traffic
jam} in the orbits of the bulge stars, when they happen to be closed to
the center.
To evaluate the possible change in the derived M/L, we will approximate
for simplicity, the
M31 central region as exactly edge-on (with the $Ox$ axis pointing
towards us), axisymmetric, and cylindrically rotating around the axis
$Oz$.

The second order hydrodynamic (HD) equation in $x$ is then:
\begin{equation}
\frac{\partial\left(\nu \overline{V_x^2}\right)}{\partial x}
+ \frac{\partial\left(\nu \overline{V_x V_y}\right)}{\partial y}
= \nu \frac{\partial \Phi}{\partial x}
\end{equation}
where $\Phi$ is the total gravitational potential, $\nu$ the light spatial
density (assuming a constant M/L ratio), and $\overline{V_x^2}$,
$\overline{V_x V_y}$ two terms of the velocity dispersion tensor.
With the classical notations $\sigma_r^2$, $\sigma_{\theta}^2$ for the
principal axis of the velocity ellipsoid, and $\overline{\Theta}$ the
mean rotation velocity around the $Oz$ axis, we have:
\begin{equation}
\overline{V_x V_y} = \frac{x y}{x^2 + y^2} \left(\sigma_r^2 -
\sigma_{\theta}^2 - \overline{\Theta}^2 \right)
\end{equation}
Multiplying the hydrodynamical equation by $x\cdot \diff x \diff y \diff z$,
and integrating along the cylinder of axis $Ox$ surrounding the nucleus,
gives:
\begin{eqnarray}
& & \tint_{cyl.} x \frac{\partial \left(\nu \overline{V_x^2}\right)}
{\partial x} \diff x \diff y \diff z + \nonumber \\
& & \tint_{cyl.} x \frac{\partial \left(\nu \overline{V_x V_y}\right)}
{\partial y} \diff x \diff y \diff z =
\tint_{cyl.} \nu x \frac{\partial \Phi}{\partial x}
\diff x \diff y \diff z
\end{eqnarray}
Integrating by part gives a generalized local form of the vector
virial theorem, as:
\begin{eqnarray}
W_{xx} & = & \tint_{cyl.} \nu x \left| \frac{\partial \Phi}{\partial x}
\right| \diff x \diff y \diff z \nonumber \\
\vphantom{W_{xx}}
& = & \mbox{L}_{\mbox{T}} \ave{\mu_x^2} + 8 \!\! \int_0^{\infty} \!\!\!
\!\! \int_0^{z_0} \frac{\nu x^2 y_0}{x^2 + y_0^2}
\left(\overline{\Theta}^2 + \sigma_{\theta}^2 - \sigma_r^2
\right) \diff x \diff z
\end{eqnarray}
$\mbox{L}_{\mbox{T}}$ is the luminosity integrated inside the cylinder.
$\ave{\mu_x^2}$
is the non-centered second order velocity moment along the x-axis,
luminosity-averaged inside the cylinder.
Apart from the familiar form of the vector virial theorem, there is
an additional term at right, which depends on the kinematical motions
in the galaxy, and which is given here as a surface integral over
one-eight of the cylinder (defined by $x, y, z > 0$).
As expected, an excess of circular over radial motions would lead
to an underestimate of the true M/L ratio when this correcting term
is not taken into account (and, of course, to an overestimate in the
opposite case).

To evaluate this correcting term, we need kinematical data for the bulge
stars around the nucleus. Many such studies have been made, e.g. by
Simien et al (1979), Mc Elroy (1983) and Kent (1989). Their data show
that the bulge is nearly isotropic, from the edge of the nucleus to at
least 5 arcmin radius. We can thus approximate
$\overline{\Theta}^2 + \sigma_{\theta}^2 - \sigma_r^2 \sim
\overline{\Theta}^2$. A fairly good approximation of the spatial
rotation curve $\overline{\Theta}^2$ is given in Monnet and Rosado (1981)
as :
\begin{equation}
\overline{\Theta}^2 = \left(720\right)^2 \frac{\conv{x}^2+\conv{y}^2}
{\left(1 + 2\left(2\sqrt{\conv{x}^2+\conv{y}^2}\right)^{3/2}\right)^2}
\end{equation}
$\overline{\Theta}^2$ is in \kms, and $\conv{x}$, $\conv{y}$ are in units
of the De Vaucouleur's equivalent radius of the bulge, $r_e$ = 1050
arcsec. The integration needed to compute the correcting term extends
at most 2 arcsec above and below the equatorial plane, and we can
safely assume that the rotation law is barotropic (i.e. independent
of $z$).

Practically speaking, $\ave{\mu_x^2}$ has been measured inside
an ellipse of $6 \times 4$ arcsec diameter, aligned along the
PA of the nucleus (see Sect. \ref{par:mu2}). Its experimental
value is $\left(185\right)^2 \mbox{km}^2\mbox{.s}^{-2}$.
The total luminosity $\mbox{L}_{\mbox{T}}$ inside the cylinder, in the I band,
has been computed using the Gaussian decomposition of the
projected intensities given in Table \ref{tab:vir}. This gives
$\mbox{L}_{\mbox{T}}$ = 1.852 \ten{6}
\Lsun$\mbox{pc}^{-2}\left(\arcsec\right)
^{-1}$. It includes, of course, the contributions from both the nucleus
and the bulge components. The correcting term is then:
\begin{equation}
C  =  8 \int_0^{\infty} \!\! \int_0^{2\arcsec}
\frac{\nu x^2 y_0 \overline{\Theta}^2}{x^2 + y_0^2} \diff x \diff y
 =  4.0 10^{8} \mbox{L}_{\sun}\mbox{km}^2\mbox{.s} ^{-2}
\end{equation}
to be compared with:
\begin{equation}
\mbox{L}_{\mbox{T}} \ave{\mu_x^2} =
6.4 10^{10} \mbox{L}_{\sun}\mbox{km}^2\mbox{.s} ^{-2}
\end{equation}
the correction factor is only 0.6\%, and we can safely
treat the nucleus as isolated.

\section{Appendix: Virial mass of a tumbling bar}
\label{par:app2}

Bar's tumbling motions will change the virial theorem and,
for a favourable orientation, the mass-to-light ratio ($\Gamma$)
can be lowered. We evaluate the maximum possible effect below:

As shown by Binney (1980), the effect of the rotation of the figure of
a triaxial galaxy, with a constant angular velocity $\Omega$ around
its $z$ axis, is to add an effective kinetic
energy $\Omega^2 \left( {\cal I}_{xx} - {\cal I}_{yy} \right)$ in the
direction of the longer axis $x$ in the equatorial plane.
The same energy is of course removed in the direction of the shorter axis
$y$ in the equatorial plane. In that formula,
${\cal I}_{xx}$ and ${\cal I}_{yy}$ are defined as:
\begin{equation}
{\cal I}_{xx} = \int_{{\bbbr}^3} \rho x^2 \; \diff x \diff y \diff z
\;;\;\;\; {\cal I}_{yy} = \int_{{\bbbr}^3} \rho y^2 \; \diff x \diff y
\diff z
\end{equation}
With a mass density ($\rho$):
\begin{equation}
\rho \left( x,y,z \right)  =  \Gamma \sum_i I_i
\exp{ \left\{- \frac{1}{ 2 t_i u_i \sigma_i^2}
\left( t_i^2 x^2 + u_i^2 y ^2 +
z^2 \right) \right\} }
\end{equation}
direct integration gives:
\begin{eqnarray}
{\cal I}_{xx} & = & \left( 2 \pi \right)^{3/2} \Gamma \sum_i \frac{\left(
u_i \right)^{3/2}}
{\left( t_i \right)^{1/2}} I_i \sigma_j^5 \\
{\cal I}_{yy} & = & \left( 2 \pi \right)^{3/2} \Gamma \sum_i \frac{\left(
t_i \right)^{3/2}}
{\left( u_i \right)^{1/2}} I_i \sigma_i^5
\end{eqnarray}
The effective energy $\delta {\cal E}$, added to the virial energy on the
$x$ axis, is then:
\begin{equation}
\delta {\cal E} = \Omega^2 \left( {\cal I}_{xx} - {\cal I}_{yy} \right) =
\left( 2 \pi \right)^{3/2} \Gamma \Omega^2 \sum_i I_i \frac{\left( u_i^2 -
t_i^2 \right)} {\left( t_i u_i \right)^{1/2}} \sigma_i^5
\end{equation}
to be compared with the virial energy:
\begin{equation}
\mbox{M}_{\mbox{T}} \ave{\mu^2} = \ave{\mu^2} \sum_i \mbox{M}_i =
\left( 2 \pi \right)^{3/2} \Gamma \ave{\mu^2}  \sum_i \left( t_i u_i
\right)^{1/2} \sigma_i^3 I_i
\end{equation}
Their ratio ($\lambda$) is:
\begin{equation}
\lambda = \frac{\Omega^2 \left( {\cal I}_{xx} - {\cal I}_{yy} \right)}
{\mbox{M}_{\mbox{T}} \ave{\mu^2}}
\end{equation}
which gives:
\begin{equation}
\lambda = \frac{\Omega^2}{\ave{\mu^2}} \frac{ \displaystyle{
\sum_i \left(t_i u_i\right)^{-1/2} \left(u_i^2 - t_i^2\right)
\sigma_i^5 I_i} }
{\displaystyle{ \sum_i\left(t_i u_i\right)^{1/2} \sigma_i^3 I_i} }
\end{equation}
Note that $\lambda$ is an adimensional parameter.

We will assume that the nucleus of M31 is viewed along its longest axis Oz,
which gives the maximum possible decrease of the M/L ratio $\Gamma$,
namely $\Gamma = \left(1 - \lambda\right) \Gamma_0$, where $\Gamma_0$
is the precedently mass-to-light ratio computed with the Virial theorem,
assuming no tumbling motion.

Note that the nucleus rotates around its smallest axis $\mbox{O}_z$, and
that the predicted velocity field will mimic that of an axisymmetric
galaxy, in agreement with the experimental results.
{}From Monnet et al. (1992) formulae for the projection of a triaxial
galaxy light distribution, and with the Euler angles $\theta = \pi/2$,
$\phi = \pi/2$, $\psi = 0$, the relationships between the spatial
parameters $t_i, u_i, \sigma_i, I_j$ of the Gaussians which
describe the nucleus light intensity law and their projected
counterparts $\proj{q}_i, \proj{\sigma}_i, \proj{I}_i$
(as given in Table \ref{tab:vir}) are:
\begin{equation}
\proj{q}_i = u_i \; ; \;  \proj{\sigma}_i^2 = \frac{t_i}{u_i}
\sigma_i^2
\; ; \; \proj{I}_j = \left(2 \pi \frac{u_i}{t_i}\right)^{1/2}
\sigma_i I_j
\end{equation}
There is only one free parameter $t_i$, not constrained by the
experimental data. For simplicity, we take $k = u_i / t_i$,
independant of $i$  ($k > 1$).
This gives:
\begin{equation}
\lambda = \frac{\Omega^2}{\ave{\mu^2}} \left(k^2 - 1\right)
\frac{\displaystyle{\sum_i \proj{q}_i \proj{\sigma}_i^4
\proj{I}_i} }
{\displaystyle{\sum_j \proj{q}_i \proj{\sigma}_i^2 \proj{I}_i} }
\end{equation}
{}From Monnet et al (1992) Appendix A, the spatial angular velocity
$\Omega$ of the bar and its projected value $\overline{\Omega}$ are
then easily related by: $\overline{\Omega} = k \Omega$.
Finally we obtain:
\begin{equation}
\lambda = \frac{k^2 - 1}{k^2} \frac{\overline{\Omega}^2
\overline{\sigma}^2} {\ave{\mu^2}}
\end{equation}
where $\overline{\sigma}$ is a characteristic length, given by:
\begin{equation}
\overline{\sigma}^2 = \frac{\displaystyle{\sum_i \proj{q}_i
\proj{\sigma}_i^4 \proj{I}_i} }
{\displaystyle{\sum_i \proj{q}_i \proj{\sigma}_i^2 \proj{I}_i} }
\end{equation}

Using the model E (Table \ref{tab:vir}) gives
$\overline{\sigma} = 1.32$ arcsec.
A reasonable upper limit for $\overline{\Omega}$ is obtained by taking
the corotation radius of the bar at the edge of the nucleus, i.e. at
a radius of about 2 arcsec, where the deconvolved (but not deprojected)
radial velocity is $\sim 100 \mbox{km.s}^{-1}$. This gives
$\overline{\Omega}
\overline{\sigma} = 66 \mbox{km.s}^{-1}$.
With a mean non-centered second
order radial momemtum $\ave{\mu}$ = 250 \kms, the
asymptotic value for $\lambda$, attained for a highly flattened bar
($k$ large) is:
\begin{equation}
\lambda_{\infty} = \frac{\overline{\Omega}^2 \overline{\sigma}^2}
{\ave{\mu}^2} = 0.06
\end{equation}

This indeed decreases the mass-to-light ratio, but only by 16\%, even
for this highly contrived model. Thus the large values found by the
virial theorem are not substantially changed by the tumbling bar
hypothesis, and can be considered secure.

\section{Appendix: Self-gravitating dynamical models}
\label{par:dynmod}

We give here a brief description of the way we derived
axisymmetric dynamical
models using the formalism described in Emsellem et al. (1993).

The spatial luminosity distribution is given by its gaussian
expansion (See Eq. \ref{eq:nuspa}), and assuming
a constant mass-to-light ratio $\Gamma$, we thus obtain the
density distribution as ($t_i = u_i; \; q^2_i = t_i \cdot u_i$):
\begin{equation}
\rho \left( R, z \right) = \sum_i P_i \times
\exp{\left\{ - \frac{1}{2 \sigma_i^2} \left(R^2 + \frac{z^2}{q_i^2}
\right) \right\}}
\end{equation}
where $P_i = \Gamma \times I_i$.

The classical set of Jeans equations are then used to derived the
spatial dynamical quantities as follows:
\begin{equation}
\rho \mbox{\boldmath$\sigma_z$}^2 = \int_z^\infty \rho \frac{\partial \Phi}
{\partial z} \cdot \mbox{d} z
\end{equation}
and,
\begin{eqnarray}
\rho \left(\mbox{\boldmath$\sigma_{\theta}$}^2 +
\overline{\mbox{\boldmath$\Theta$}
}^2 \right)  & = & \kappa_R \cdot \left[R \frac{\partial
\left( \rho \mbox{\boldmath$
\sigma_z$}^2 \right)} {\partial R} + \left(1 + \lambda_R\right) \rho
\mbox{\boldmath$\sigma_z$}^2 \right] \nonumber \\
&& + \rho R \frac{ \partial \Phi}{\partial R}  \label{Jeantheta}
\end{eqnarray}
where we describe the radial and tangential anisotropies
by respectively $\kappa_R$ and $\kappa_{\theta}$:
\begin{equation}
\kappa_R(R,z) = \frac{\mbox{\boldmath$\sigma_R$}^2(R,z)}{\mbox{\boldmath$
\sigma_z$}^2(R,z)} \;\; \mbox{and}
\; \kappa_{\theta}(R,z) = \frac{\mbox{\boldmath$\sigma_\theta$}^2(R,z)}
{\mbox{\boldmath$\sigma_z$}^2(R,z)} \label{kappas}
\end{equation}
$\lambda_R$ being given as:
\begin{equation}
\lambda_R(R,z) = \frac{\partial\ln{\left(1 - \kappa_R(R,z)\right)}}
{\partial\ln{R}}
\end{equation}

The line of sight $Oz'$ is defined by its inclination angle $i$ with respect
to $Oz$, $(Ox',Oy')$ being the sky plane (face-on: $i=0\;$,
edge-on: $i=90 \degr$). The new cartesian coordinate system $(x', y', z')$
is then related to $(x, y, z)$ by:
\begin{equation}
\left\{
\begin{array}{ll}
x = x' \\
y = - y' \cos{i} + z' \sin{i} \\
z = y' \sin{i} + z' \cos{i}
\end{array}
\right.
\end{equation}

The derivations of the projected non-centered first and second order
moments are then readily achieved (using the same notations as in Emsellem
et al. 1993):
\begin{eqnarray}
\mu_{1}(x',y') & = & - \frac{1}{\nu_p} \int_{-\infty}^{\infty} \nu
\overline{\mbox{\boldmath$\Theta$}} \sin{i} \cos{\theta} \cdot \diff z' \\
& = & - \sin{i} \int_{-\infty}^{\infty} \cos{\theta}
\frac{\nu}{\rho^{1/2}} \nonumber \\
&& \times \left[ \rho \left( \mbox{\boldmath$\sigma_{\theta}$}
^2 + \overline{\mbox{\boldmath$\Theta$}}^2 \right)  - \kappa_{\theta} \rho
\mbox{\boldmath$\sigma_z$}^2 \right]^{1/2} \cdot \diff z' \\
\mu_{2}^2(x',y') & = & \frac{1}{\nu_p} \int_{-\infty}^{\infty} \nu \biggl(
\mbox{\boldmath$\sigma_z$}^2 \cos^2{i} + \mbox{\boldmath$\sigma_R$}^2 \sin^2
{\theta} \sin^2{i} \nonumber \\
&& + \left( \mbox{\boldmath$\sigma_{\theta}$}^2 + \overline{\mbox{
\boldmath$\Theta$}}^2 \right) \cos^2{\theta} \sin^2{i} \biggr) \cdot
\diff z' \label{2mom}
\end{eqnarray}
Hence the centered second order moment can be written as:
\begin{equation}
\varsigma_p^2(x',y') = \mu_{2}^2(x',y') - \mu_{1}^2(x',y')
\end{equation}

This leads to a double integration for each moment,
except in the isotropic
case for which the projected second order moment is obtained through
a single integration (see Emsellem et al. 1993).
The convolved projected velocity and velocity fields (respectively
$\conv{V}(x',y')$ and $\conv{\mbox{\boldmath$\sigma$}}(x',y')$
are then simply given by:
\begin{eqnarray}
\overline{\nu}_p & = & \nu_p \otimes PSF \\
\conv{V} & = & \frac{\left( \nu_p \cdot \mu_{1} \right) \otimes PSF}
{\overline{\nu}_p} \\
\overline{\mu}_{2}^2 & = & \frac{\left( \nu_p \cdot \mu_{2}^2 \right)
\otimes PSF}{\overline{\nu}_p} \\
\mbox{and \hspace{1cm}}\; \conv{\mbox{\boldmath$\sigma$}}^2 & =
& \overline{\mu}_{2}^2 - \conv{V}^2
\end{eqnarray}

In the isotropic case, we have $\kappa_{\theta}$ and $\kappa_{R}$ equal to
unity. For the tangentially anisotropic case tackled in the present paper,
we used a mass weighted $\kappa_{\theta}$ anisotropy function, so that
the bulge remains isotropic while the nucleus is highly anisotropic:
\begin{equation}
\kappa_{\theta}(R,z) = \frac{\alpha \cdot \left(\rho_N\right)^{\beta} +
\left(\rho_B\right)^{\beta}}{\left(\rho\right)^{\beta}}
\end{equation}
where $\alpha$ and $\beta$ are two free parameters fixed by the user
($\alpha$ and $\beta$ fixed to 1 leads to the simple isotropic case),
$\rho_N$ and $\rho_B$ being respectively the density distribution
of the nucleus and the bulge
(derived from their respective gaussian components).


\begin{thebibliography}{}
\bibitem{} Bacon R., Adam G., Baranne A., Courtes G., Dubet D., Dubois
J.P., Georgelin Y., Monnet G., Pecontal E., Urios J., 1988,
in: VLTs, their Instrumentation V.II, ESO, Conf.
Ulrich M.-H. (ed.), Garching, 21-24 March 1988, p. 1185
\bibitem{} Bacon R., Georgelin Y., Monnet G., 1990, CFH bull., 23, p??
\bibitem{} Binney J., 1980, in: The Structure, Evolution of Normal
Galaxies, p. 57, ed. S. Fall, D. Lynden-Bell, Cambridge University
Press
\bibitem{} Binney J., Tremaine S., 1987, in: Galactics Dynamics, Princeton
series in Astrophysics, Ostriker J. ed.
\bibitem{} Bontz R.J., 1979, ApJ, 233, 402
\bibitem{} Bender R., 1990,  A\&A, 229, 441
\bibitem{} Bender R., 1993, in: ESO/EIPC workshp, "Structure, dynamics
and chemical evolution of elliptical galaxies", p. 3, ed. Danziger J., Isola
d'Elba, May 1992
\bibitem{} Courtes G., Georgelin Y., Bacon R., Monnet G., Boulesteix J.,
1987, in: Santa Cruz Summer Workshop, "Instrumentation for ground based
optical astronomy", July 1987
\bibitem{} Crane P. and Stiavelli M., 1993,
in: ESO/EIPC workshp, "Structure, dynamics
and chemical evolution of elliptical galaxies", p. 137, ed. Danziger J., Isola
d'Elba, May 1992
\bibitem{} Dressler A., Richstone D.O., 1988, ApJ, 324, 701 (DR88)
\bibitem{} Emsellem E., Monnet G., Bacon R., 1993a, A\&A, submitted
\bibitem{} Emsellem E., Monnet G., Bacon R., Nieto J.L., 1993b, A\&A,
submitted
\bibitem{} Gerhard O.E., 1986, MNRAS, 219, 373
\bibitem{} Gerhard O.E., 1988, MNRAS, 232, 13
\bibitem{} Harris W.E., Racine R., 1979, ARA\&A, 17, 241
\bibitem{} Kent S.M., 1989, AJ, 97, 1614
\bibitem{} Kormendy J., 1988, ApJ, 325, 128 (K88)
\bibitem{} Lallemand A., Duschene M., Walker M.F., 1960, PASP, 72, 76.
\bibitem{} Lauer T.R., Faber S.M., Groth E.J., Shaya E.J., Campbell B.,
Code A., Currie D.G., Baum W.A., Ewald S.P., Hester J.J., Holtzman J.A.,
Kristian J., Light R.M., Lynds C.R., O'Neil E.J., Westphal J.A., 1993,
NOAO preprint 521 (AJ submitted)
\bibitem{} Light E.S., Danielson, R.E., Schwarzschild, M, 1974, ApJ, 194, 257
\bibitem{} McElroy D.B., 1983, ApJ, 270, 485
\bibitem{} McLure R., Grundmann W.A., Rambold W.N., Fletcher J.M.,
Richardson J.M., Stillburn E.H., Racine J.R., Christian C.A., Waddel P.,
1989, PASP, 101, 1156
\bibitem{} McLure R., Arnaud J., Fletcher J.M., Nieto J.L., Racine J.R.,
1991, PASP, 103, 570
\bibitem{} Miller R.H., Smith B.F., 1992, ApJ, 393, 508
\bibitem{} Monnet G., Bacon R., Emsellem E., 1992, A\&A, 253, 366
\bibitem{} Monnet G., Rosado M., 1981, A\&A, 102, 175
\bibitem{} Mould J., Graham J., Matthews K., Soifer B.T., Phinney E.
S., 1989, ApJ 339, L21
\bibitem{} Nieto J.L., Macchetto F.D., Perryman M.A.C., Serego
Alighieri S., Lelievre G., 1986, A\&A 165, 189.
\bibitem{} Nieto J.L., 1984, ApJ, 287, 108
\bibitem{} Peterson C.J., King I.R., 1975, AJ, 80, 427
\bibitem{} Richstone D., Bower G., Dressler A., 1990, ApJ, 353, 118
\bibitem{} Rousset A., Pecontal E., Bacon R., in: Acta Stereologica,
Proceedings of the eighth international congress for stereologie, II,
February 1992, p593.
\bibitem{} Sargent W.L.W, Schechter P.L., Boksenberg A., Shortridge K.,
1977, ApJ, 212, 326
\bibitem{} Simien F., Pellet A., Monnet G., 1979, A\&A 72, 12
\bibitem{} Soifer B.T., Rice W.L., Mould J.R., Gillet F.C., Rowan,
Robinson M., Haring H.J., 1986, ApJ, 304, 651
\bibitem{} Stark A., 1977, ApJ, 213, 368
\bibitem{} Tonry J., Davis M., 1979, AJ, 84, 1511
\bibitem{} Wirth A., Smarr L.L., Bruno T.L., 1985, ApJ, 290, 140
\end{thebibliography}
\end{document}